\documentclass[a4paper,11pt]{article}
\usepackage{jcappub}
\usepackage{aas_macros}
\usepackage{graphicx}
\usepackage{amsmath}
\usepackage{amssymb}
\usepackage{hyperref}
\usepackage{physics}
\usepackage[capitalize]{cleveref}

\usepackage{mathrsfs}

\makeatletter
\gdef\@fpheader{\mbox{}}
\makeatother

\newlength{\fullw}
\newlength{\halfw}
\newlength{\threefigw}
\newlength{\twofigw}
\newlength{\onefigw}
\newlength{\bigfigw}
\DeclareMathOperator{\erfc}{erfc}

\newcommand{\dirac}[1]{\delta\negthinspace\left(#1\right)}

\newcommand{\heaviside}[1]{\mathrm{\Theta}\!\left( #1 \right)}

\newcommand{\boldmathsymbol}[1]{{\ensuremath{\boldsymbol{#1}}}}

\newcommand{\tetration}{\mathbin{\uparrow\uparrow}}

\newcommand{\sss}[1]{{\scriptscriptstyle{#1}}}
\newcommand{\efolds}{$e$-folds}
\newcommand{\efold}{$e$-fold}

\newcommand{\uPl}{\mathrm{Pl}}
\newcommand{\uend}{\mathrm{end}}
\newcommand{\uinf}{\mathrm{inf}}
\newcommand{\uqw}{\mathrm{qw}}
\newcommand{\ubw}{\mathrm{bw}}

\newcommand{\uSI}{\mathrm{SI}}
\newcommand{\uLT}{\mathrm{LT}}

\newcommand{\usssPl}{\sss{\uPl}}

\newcommand{\usssSI}{\sss{\uSI}}
\newcommand{\usssLT}{\sss{\uLT}}
\newcommand{\umode}{\mathrm{mode}}
\newcommand{\zero}{{_0}}

\newcommand{\Mp}{M_\usssPl}
\newcommand{\Mpc}{\mathrm{Mpc}}
\newcommand{\Gpc}{\mathrm{Gpc}}

\newcommand{\Nqw}{N_\uqw}

\newcommand{\Nzero}{N_\zero}
\newcommand{\Nmax}{N_{\max}}

\newcommand{\deltaN}{\var N}

\newcommand{\rvp}[1]{\bar{#1}}
\newcommand{\rv}[1]{\Delta#1}
\newcommand{\Pname}{P}
\newcommand{\Pof}[1]{\Pname\negthinspace\left(#1\right)}
\newcommand{\Pini}[1]{\Pname_\zero\negthinspace\left(#1\right)}
\newcommand{\Pjt}[1]{\Pof{#1}}
\newcommand{\Pfw}[1]{\Pof{#1}}
\newcommand{\Prv}[1]{\rvp{\Pname}\negthinspace\left(#1\right)}
\newcommand{\Plt}[1]{\Pname_{\usssLT}\negthinspace\left(#1\right)}
\newcommand{\Pbw}[1]{\Pname_{\ubw}\negthinspace\left(#1\right)}

\newcommand{\Xrv}{\rvp{X}}
\newcommand{\Frv}{\rvp{F}}
\newcommand{\Grv}{\rvp{G}}

\newcommand{\Frvzero}{\Frv_{\zero}}
\newcommand{\Frvinfty}{\Frv_{\infty}}

\newcommand{\Hinf}{H_\uinf}
\newcommand{\Ginf}{G}

\newcommand{\phiend}{\phi_\uend}
\newcommand{\phiqw}{\phi_\uqw}

\newcommand{\phistar}{\phi_*}
\newcommand{\phizero}{\phi_\zero}
\newcommand{\phimax}{\phi_{\max}}

\newcommand{\chizero}{\chi_\zero}
\newcommand{\chihat}{\hat{\chi}}
\newcommand{\chizerohat}{\chihat_\zero}

\newcommand{\tzero}{t_\zero}
\newcommand{\xzero}{x_\zero}

\newcommand{\kstar}{k_*}

\newcommand{\epsV}[1]{\epsilon_{v_{#1}}}
\newcommand{\etaSI}{\eta_\usssSI}

\newcommand{\bx}{\boldmathsymbol{x}}

\newcommand{\zetahat}{\hat{\zeta}}

\begin{document}

\title{Time-reversed Stochastic Inflation}

\author{Baptiste Blachier}
\author{and Christophe Ringeval}
\affiliation{Cosmology, Universe and Relativity at Louvain (CURL),
Institute of Mathematics and Physics, University of Louvain,
2 Chemin du Cyclotron, 1348 Louvain-la-Neuve, Belgium}

\emailAdd{baptiste.blachier@uclouvain.be}
\emailAdd{christophe.ringeval@uclouvain.be}

\date{\today}

\abstract{Cosmic inflation may exhibit stochastic periods during which
  quantum fluctuations dominate over the semi-classical
  evolution. Extracting observables in these regimes is a notoriously
  difficult program as quantum randomness makes them fully
  probabilistic. However, among all the possible quantum histories,
  the ones which are relevant for Cosmology are conditioned by the
  requirement that stochastic inflation ended. From an observational
  point of view, it would be more convenient to model stochastic
  periods as starting from the time at which they ended and evolving
  backwards in times.

  We present a time-reversed approach to stochastic inflation, based
  on a reverse Fokker-Planck equation, which allows us to derive
  non-perturbatively the probability distribution of the field values
  at a given time before the end of the quantum regime. As a motivated
  example, we solve the flat semi-infinite potential and derive a new
  and exact formula for the probability distribution of the
  quantum-generated curvature fluctuations. It is normalisable while
  exhibiting tails slowly decaying as a Levy distribution. Our
  reverse-time stochastic formalism could be applied to any
  inflationary potentials and quantum diffusion eras, including the
  ones that can lead to the formation of primordial black holes.}


\maketitle

\section{Introduction}
\label{sec:intro}

Cosmic inflation is an hypothesised period of accelerated expansion of
the universe which has taken place in its earlier instants. The
kinematics of accelerated space-time solves at once the so-called
flatness and horizon problems that plague the standard Hot Big-Bang
model of Friedmann and Lema\^itre~\cite{Starobinsky:1979ty,
  Starobinsky:1980te, Guth:1980zm, Linde:1981mu, Albrecht:1982wi,
  Linde:1983gd, Mukhanov:1981xt, Mukhanov:1982nu, Starobinsky:1982ee,
  Guth:1982ec, Hawking:1982cz, Bardeen:1983qw}. Inflation can be
triggered in many ways~\cite{DeFelice:2011uc, Baumann:2014nda,
  BeltranJimenez:2015xxv, Vennin:2015vfa}, but the simplest of all
consists in a self-gravitating scalar
field~\cite{Mukhanov:1990me}. Cosmic inflation generated by a single
scalar field slow-rolling along its potential comes with a set of
additional observable predictions. The small quantum fluctuations of
the field-metric system give rise to super-Hubble curvature
fluctuations and primordial gravitational waves which, later on, act
as the seeds of the Cosmic Microwave Background (CMB) anisotropies and
large scale structures (LSS) of the universe. The power spectra of both the
scalar and tensor fluctuations can be rigorously calculated by
semi-classical methods and are compatible with the cosmological
measurements~\cite{Mukhanov:1987pv, Mukhanov:1988jd, Stewart:1993bc,
  Liddle:1994dx, Nakamura:1996da, Gong:2001he, Hoffman:2000ue,
  Schwarz:2001vv, Leach:2002ar, Schwarz:2004tz, Casadio:2005xv,
  Martin:2013uma, BeltranJimenez:2013ikr, Auclair:2022yxs,
  Bianchi:2024qyp}. The quantum origin of these perturbations implies
a very small amount of primordial non-Gaussianities while the
single-field dynamics enforces that all cosmological perturbations are
adiabatic; both predictions being again compatible with present
observations~\cite{Planck:2018jri, Planck:2019kim,
  Chowdhury:2019otk}. Narrowing down single-field slow-roll inflation
to some specific models, each of them being defined by the shape of
its potential, makes more accurate predictions~\cite{EIopiparous}. As
shown in \cite{Martin:2024nlo}, the running of the spectral index
should be small and negative, a prediction that will be soon put to
the test with the Euclid satellite data~\cite{Euclid:2021qvm,
  Euclid:2024yrr, Euclid:2024ris}. As of today, the single-field
models favoured by the data have plateau-like potentials, at least
within the domain where cosmological perturbations seeding CMB and LSS
are generated~\cite{Martin:2013nzq, Martin:2024qnn}.

\begin{figure}
\begin{center}
  \includegraphics[width=\onefigw]{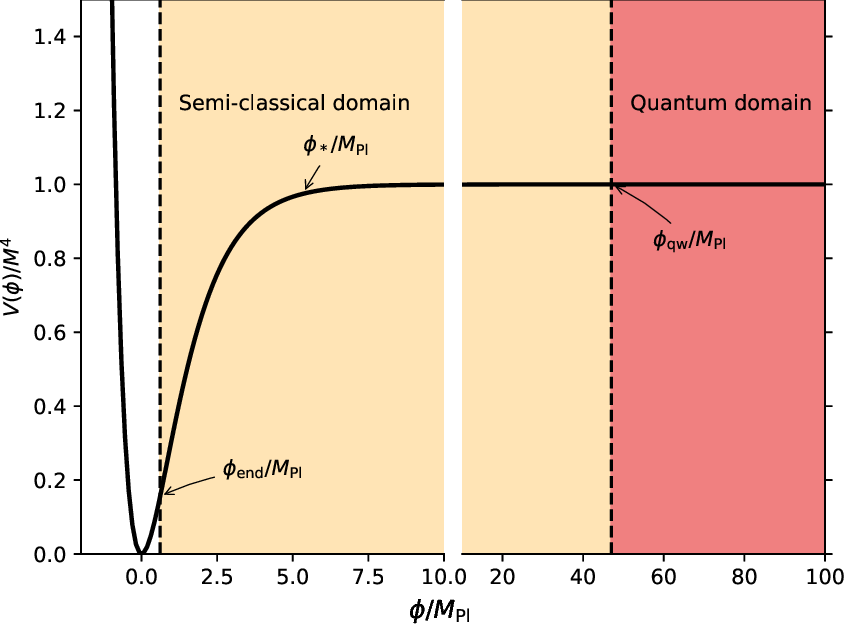}
\caption{Normalised potential for the Starobinsky model in the
  Einstein frame as a function of the field values $\phi/\Mp$. In the
  semi-classical domain, for $\phi\in[\phiend,\phiqw]$, the universe
  is inflating while the field rolls down the potential. This is the
  regime in which cosmological perturbations are generated, around
  $\phi=\phistar$: quantum fluctuations are dealt with over a
  classical background. For $\phi>\phiqw$, quantum fluctuations become
  of larger amplitude than the classical field evolution, the
  semi-classical treatment breaks down: this is the regime of
  stochastic inflation. Because the quantum domain is at field values
  very far from $\phiend$, any structure formed in this region would
  be of length scales much larger than the Hubble radius today and, as
  such, can only affect background cosmological quantities.}
\label{fig:sidomains}
\end{center}
\end{figure}

A proto-typical example of plateau inflation is given by the
Starobinsky model~\cite{Starobinsky:1980te}, the potential of which,
in the Einstein frame, reads
\begin{equation}
V(\phi) = M^4 \left(1- e^{-\sqrt{\frac{2}{3}} \frac{\phi}{\Mp}} \right)^2.
\label{eq:sipot}
\end{equation}
This potential is represented in \cref{fig:sidomains}. Up to some very
small corrections, this is also the potential of the so-called Higgs
inflation model~\cite{Bezrukov:2007ep}.  In these scenarios, cosmic
inflation occurs while the field slowly rolls down the potential, from
large to small field values, in a regime where the potential dominates
over the kinetic energy. Progressively the field accelerates, up to a
field value $\phiend$ at which the kinetic energy overcomes the
potential energy and inflation stops~\cite{Auclair:2024udj}. The
cosmological perturbations seeding the large scale structures of the
universe are generated in a small region around $\phi=\phistar$, the
field value at which a wavelength of astrophysical interest today
(typically $\kstar=0.05\,\Mpc^{-1}$) crossed the Hubble radius during
inflation. For Starobinsky inflation, a typical value is $\phistar
\simeq 5 \Mp$~\cite{EIopiparous}. The domain $\phi<\phiend$
corresponds to the post-inflationary phase, the reheating epoch,
during which the universe transits from being field-dominated to the
standard hot Big-Bang radiation era~\cite{Kofman:1997yn,
  Bassett:2005xm, Martin:2006rs, GarciaBellido:2008ab, Martin:2010kz,
  Demozzi:2012wh, Martin:2014nya, Terada:2014uia, Martin:2016oyk,
  Drewes:2019rxn, Joana:2022uwc}.

As represented in this figure, there is also an upper limit for the
field value, $\phiqw$, above which the potential is so flat that
quantum fluctuations dominate over the classical drift induced by the
tilt of the potential. In this domain, semi-classical calculations for
both the field evolution and its perturbations break down and we will
refer to this part of the potential as the quantum diffusion
region. In order to determine the field value $\phiqw$, one can use
the classicality criterion derived in
Refs.~\cite{Brandenberger:2015rra, Vennin:2015hra}. Defining the dimensionless function
\begin{equation}
\eta(\phi) = \dfrac{V(\phi)}{24 \pi^2 \Mp^4} \left|1 + \dfrac{\epsV{2}(\phi)}{4
  \epsV{1}(\phi)} \right|,
\end{equation}
where~\cite{Liddle:2003py,Vennin:2014xta}
\begin{equation}
\epsV{1}(\phi)  = \dfrac{\Mp^2}{2} \left(\dv{\ln V}{\phi}\right)^2,\quad
\epsV{2}(\phi)  \equiv \sqrt{2 \epsV{1}} \Mp \, \dv{\ln
  \qty|\epsV{1}|}{\phi} = 2 \Mp^2 \left[\left(\dv{\ln
    V}{\phi}\right)^2 - \dfrac{1}{V} \dv[2]{V}{\phi} \right],
\label{eq:epsVdef}
\end{equation}
the classical drift dominates over quantum fluctuations provided
$\eta(\phi) \ll 1$. We can therefore estimate the location of the
quantum wall separating the classical region from the quantum region
by solving $\eta(\phiqw)=1$. For Starobinsky inflation, using
\cref{eq:sipot}, one gets
\begin{equation}
\etaSI(\phiqw) = \dfrac{M^4}{48 \pi^2 \Mp^4} e^{-2\sqrt{\frac{2}{3}}
  \frac{\phiqw}{\Mp}} \left(e^{\sqrt{\frac{2}{3}}\frac{\phiqw}{\Mp}}-1 \right)^2
\left(e^{\sqrt{\frac{2}{3}}\frac{\phiqw}{\Mp}} + 2 \right) = 1.
\end{equation}
There is an exact, but not particularly illuminating, solution to this
equation, which can be approximated, at large field values, by
\begin{equation}
\phiqw \simeq \sqrt{\dfrac{3}{2}} \Mp \ln\left(\dfrac{48 \pi^2
  \Mp^4}{M^4}\right) \simeq 47 \Mp.
\end{equation}
Here, we have used the value $M\simeq 3.3 \times 10^{-3}\Mp$ which is
required to match the amplitude of the CMB
anisotropies~\cite{Martin:2024qnn}. An order of magnitude of the
maximal number of {\efolds} that could be made in the semi-classical
domain, for $\phi \in [\phiend,\phiqw]$, ends up being
$\order{10^{16}}$. This figure is, as expected, much larger than the
$\order{100}$ {\efolds} required to solve the problems of the Big-Bang
model and that is why, in all practical applications, the field
dynamics coming from the quantum region can be safely
ignored\footnote{Notice that the value of $\phiqw$ depends on the
inflationary potential and so does the maximal classical number of
{\efolds}.}. Let us mention that there exist inflationary models for
which the quantum region is not present at all~\cite{Mukhanov:2014uwa,
  Dimopoulos:2016yep}.

Nonetheless, all potentials having an infinitely flat plateau possess
a quantum diffusion domain at large field values. Since the early days
of the inflationary paradigm, it has been realised that the quantum
diffusion regime can produce a very different spacetime than the
smooth quasi de Sitter universe associated with the semi-classical
domain. For instance, inflation can become
eternal~\cite{Vilenkin:1983xq, 1986PhLB..175..395L,Goncharov:1987ir}
possibly generating a multiverse structure and its associated
divergences and measure problems~\cite{Winitzki:2006rn,
  Winitzki:2008yb, Martin:2016iqo, Tomberg:2025fku}.

Curvature fluctuations created in the quantum domain, at
$\phi>\phiqw$, are necessarily of wavelengths much larger than the
Hubble radius today, by a factor typically given by the maximal number
of classical {\efolds}. In other words, the quantum region can affect
the structure of our universe but on length scales typically greater
than a tetration ${10\tetration3}\,\Gpc$ (for Starobinsky inflation,
this is $10^{10^{16}}\,\Gpc$). Although effects coming from the
quantum diffusion regime may seem out of any observational reach,
background cosmological quantities, and their correlations, are
actually sensitive to super-Hubble cosmological
fluctuations~\cite{Ringeval:2010hf, Kleban:2012ph, Glavan:2017jye,
  Ringeval:2019bob, Blachier:2023ooc}. In that respect, there is an
interest in quantifying the statistics of the curvature fluctuations
created at $\phi>\phiqw$ and this is the topic of the present
paper.

In the context of primordial black holes (PBH), let us stress that
various works have been interested in a compact quantum diffusion
region, where the potential is exactly flat but only within a finite
domain of field values. This is the so-called ``quantum well'',
usually assumed to be located around the minimum of the potential,
which would then be entered after the end of semi-classical
inflation. Depending on the well width, these scenarios have been
shown to generate large enough curvature fluctuations to form
primordial black-holes~\cite{Pattison:2017mbe, Ezquiaga:2019ftu,
  Animali:2022otk, Stamou:2023vft, Stamou:2023vwz, Animali:2024jiz,
  Animali:2025pyf}. Our setup, here, is a quantum well with infinite
width, a situation known to produce divergent results for various PBH
observables~\cite{Ando:2020fjm, Tada:2021zzj}. As we will show later
on, our approach allows us to derive an exact, and finite, analytical
expression for the curvature fluctuations generated in the large well
width limit.

In practice, the evolution of a self-gravitating scalar field in the
quantum dominated region can be modelled using the stochastic
inflation formalism~\cite{Starobinsky:1986fx, Goncharov:1987ir,
  Nambu:1987ef, Kandrup:1988sc, Nakao:1988yi, Starobinsky:1994bd,
  Linde:1993xx}. It has been developed and applied to various
contexts, ranging from the study of eternal inflation to predicting
the abundances and distribution of primordial black holes. It can be
implemented in lattice simulations, binary tree statistics and efforts
are made to include it within numerical relativity
simulations~\cite{Salopek:1998qh, Fujita:2013cna, Ando:2020fjm,
  Tada:2021zzj, Tomberg:2022mkt, Mizuguchi:2024kbl, Launay:2024qsm,
  Animali:2025pyf}. Complemented with the so-called
$\deltaN$-formalism, it allows for the derivation of the statistics of
the space-time curvature fluctuations generated during quantum
diffusion~\cite{Sasaki:1995aw, Sasaki:1998ug, Wands:2000dp,
  Lyth:2004gb, Lyth:2005fi, Fujita:2014tja}. In short, the formalism
relies on coarse-graining quantum operators on sub-Hubble scales to
derive an equation of evolution for the coarse-grained field on
super-Hubble scales~\cite{Finelli:2008zg, Finelli:2010sh,
  Garbrecht:2014dca, Vennin:2020kng, Tanaka:2023gul,
  Tomberg:2024evi}. This is reminiscent of effective field theories in
which the high-energy degrees of freedom are integrated out to
understand the dynamics in the infrared. Even though the
coarse-graining erases the quantum entanglement properties of the
system, it preserves, in a non-perturbative manner, quantum
randomness. The resulting equations of motion are stochastic
differential equations and allow us to make predictions in regimes
inaccessible to the usual perturbative quantum loop expansions. For
slow-roll potentials $V(\phi)$, the quantum diffusion on the velocity
can be shown to be negligible and the dynamics of the coarse-grained
field-metric system reduce to a Langevin
equation~\cite{Lesgourgues:1996jc, Vennin:2015hra, Grain:2017dqa}
\begin{equation}
\dv{\phi}{N} = -\dfrac{1}{3 H^2} \dv{V}{\phi} + \dfrac{H}{2\pi} \xi(N).
\label{eq:langevin}
\end{equation}
Starting with this equation, we will be working from now on in Planck
units with $\Mp=1$. In \cref{eq:langevin}, the time variable is $N=\ln
a$, the number of {\efolds}, $a(t)$ being the
Friedmann-Lema\^itre-Robertson-Walker (FLRW) scale factor, $H(\phi)$ is
the Hubble parameter, $\phi$ is the coarse-grained field, and, $\xi$
is a Gaussian white noise of unit variance and vanishing mean emerging
from the coarse-graining. Starting at a given field value, say
$\phizero$ at $N=\Nzero$, the noise term will produce many possible
paths leaving, or not, the quantum diffusion domain. By construction,
their statistics is representative of the underlying quantum nature of
the system. In the semi-classical region, at $\phi<\phiqw$, the field
necessarily drifts towards smaller values and all trajectories exiting
the quantum domain never come back. As a result, the quantum wall at
$\phi=\phiqw$ acts as an absorbing boundary for the stochastic process
at $\phi>\phiqw$. The common approach to study stochastic inflation is
to follow the process forward in time, namely at increasing numbers of
{\efold} $N$, from $\Nzero$ to the lifetime $\Nqw$, a stochastic
quantity thus identified with the first passage time at
$\phiqw$~\cite{Tada:2021zzj}.

In this work, we approach the problem in the reverse way, moving back
in time, i.e., starting when the field exits the quantum domain at
$\phiqw$ all the way back to $\phizero$ at the time $\Nzero$. Let us
first emphasize that this is perfectly well-defined for the diffusion
processes described by \cref{eq:langevin}, which are a class of
continuous Markov processes. Indeed, the Markovian nature ensures that
the past and the future are independent given the present state and
Markov processes can be time-reversed~\cite{chung2005markov}.

A first motivation for doing so lies in the fact that the
time-reversal implicitly selects a subsample of all the possible
realisations of \cref{eq:langevin} for which quantum diffusion
ends. These are the realisations that are relevant for making
observable predictions in cosmology because inflation ended at some
point. In some sense, the time-reversal might also be viewed as a
choice of a measure which is discarding all never ending inflationary
realisations from the statistics. However, as we will see in
\cref{sec:rvdeltaN}, we can still include infinitely long, but still
ending, realisations of \cref{eq:langevin} and this allows us to
derive a normalisable distribution for the curvature fluctuations
encompassing all eternally inflating domains.

Another motivation for a time-reversal approach lies in the
$\deltaN$-stochastic
formalism~\cite{Fujita:2013cna,Fujita:2014tja,Vennin:2015hra}. Based
on the separate universe picture, the curvature fluctuations $\zeta$
on constant energy density hypersurface is directly related to the
fluctuations in the number of {\efolds} $\zeta = N - \ev{N}$
associated with \cref{eq:langevin}. Starting from a fixed
configuration $(\phizero,\Nzero)$, the statistics of the curvature
fluctuations for local observers are, among others, intrinsically
related to the randomness of the lifetimes $\Nqw-\Nzero$, i.e., always
in reference to the {\efold} at which inflation
ends~\cite{Tada:2016pmk}.  As we will see in \cref{sec:sols}, by
time-reversing stochastic inflation, not only the reference epoch is
necessarily the end of quantum diffusion but the reversed processes
end up being conditioned to their lifetimes and the expectation values
entering into the definition of $\zeta$ become easier to calculate.

Although we are not aware of any other work discussing time-reversed
stochastic inflation, the topic of time reversal for stochastic
processes is not new. It can be traced back to Schr\"odinger's work on
diffusion~\cite{Schrodinger1931, Chetrite:2021gcc} and to the study of
Markov processes by Kolmogorov~\cite{Kolmogoroff1936ZurTD,
  Kolmogoroff1937}. The topic was put on some mathematical ground in
the late fifties, in the fields of probability theory and mathematical
physics~\cite{Nelson1958, Meyer1962, Nagasawa1961, Nagasawa1963}. In
particular, it was realised that to obtain stationary transition
probabilities for the reverse process (i.e. enforcing temporal
homogeneity), one must reverse from a random time, not a fixed time,
leading to the foundations of a modern path-by-path theory of
reversal~\cite{Hunt1960}. The theory was later generalised to a wider
class of random times, called co-optional times in
Refs.~\cite{Chung1962, Ikeda1964, Nagasawa1964}, some results being
summarized in Refs.~\cite{Weil1967, Cartier1968, meyer1968processus,
  blumenthal2007markov}. Latest works notably revolved around
determining under which conditions the strong Markov property are
preserved in reverse processes~\cite{Chung1969ToRA}, see
Refs.~\cite{chung2005markov, Meyer1971, Rogers_Williams_2000} for
reviews.

In the following, we will consider a semi-infinite quantum diffusion
region described by \cref{eq:langevin} in which the potential remains
constant at $\phi>\phiqw$. As such, we are concerned with the
time-reversal of a simple Brownian motion, which is still a topic of
active researches in probability theory~\cite{Azma1973ThorieGD,
  Lindquist1979, Castanon1982, Anderson1982, Follmer1985, Pardoux1985,
  Haussmann1985, Haussmann1986, Elliott1985, Follmer1986,Getoor1979,
  Sharpe1980, nagasawa1993schroedinger, Revuz2004, Cattiaux2023,
  Conforti2022}. Our paper is organised as follows. In
\cref{sec:trfk}, we detail how to derive, and interpret, a
time-reversed Fokker-Planck equation associated with
\cref{eq:langevin}. We also provide an exact analytical solution to
the probability distribution of the field in the semi-infinite flat
quantum region. In \cref{sec:rvdeltaN}, we introduce and use a
time-reversed $\deltaN$-formalism to derive exact expressions for the
probability density of the curvature fluctuations. We conclude in
\cref{sec:conc} and discuss how the time-reversed stochastic inflation
formalism could be applied to more complex situations as well as to
the formation of primordial black holes.

\section{Time reversal of diffusion processes}
\label{sec:trfk}

In the quantum diffusion region, the coarse-grained field $\phi$
evolves according to \cref{eq:langevin} which describes a
one-dimensional diffusion process. On general grounds, denoting by
$(x,t)$ the spacetime coordinates, a one-dimensional diffusion obeys
an It\^o stochastic differential equation of the form\footnote{Due to
the inflationary origin of the stochastic noise, we exclusively use
It\^o stochastic integrals~\cite{Tomberg:2024evi}. The stochastic
differential equations would exhibit an extra term in the Stratonovich
scheme.}
\begin{equation}
    \dd{X(t)} = F\left[X(t),t \right] \dd{t} + G\left[X(t),t\right]
    \dd{W(t)}
\label{eq:itodiff}
\end{equation}
where $W(t)$ is a Wiener process, $F$ is the drift and the diffusion
amplitude is given by $G$. In the context of the Brownian motion, the
Wiener process boils down to a stochastic force generated by the
Gaussian white noise $\dd{W(t)} = \xi(t) \dd{t}$. As explicit by
comparing \cref{eq:langevin,eq:itodiff}, during stochastic inflation,
the time variable is given by the number of {\efolds} $N$, the value
of the coarse-grained field $\phi$ acts as the position $x$ of the
random walk while the potential provides the drift term. The diffusion
coefficient is $G=H/(2\pi)$ which gives the amplitude of the quantum fluctuations.

\subsection{Forward and backward Kolmogorov equations}

The transition probability density distribution $\Pfw{x,t|y,s}$ of an
It\^o process given by \cref{eq:itodiff}, where $t>s$, verifies the
equation~\cite{Sarkka_Solin_2019}
\begin{equation}
  \pdv{t} \Pfw{x,t|y,s} = - \pdv{x}\left[F(x,t) \Pfw{x,t|y,s}\right] +
  \dfrac{1}{2} \pdv[2]{x} \left[G^{2}(x,t)\Pfw{x,t|y,s}\right].
\label{eq:FKfw}
\end{equation}
The transition probability $\Pfw{x,t|y,s}$ allows us to reconstruct
the probability distribution $\Pfw{x,t}$ of finding the process at
$(x,t)$ given any initial distributions $\Pini{x}$, set at a time
$\tzero < t$, since
\begin{equation}
\Pfw{x,t} = \int \Pfw{x,t|y,\tzero} \Pini{y} \dd{y}.
\label{eq:Pfw}
\end{equation}
For this reason, \cref{eq:FKfw} is referred to as the ``forward''
Kolmogorov equation but we will use the more common (in the physics
community) denomination of Fokker-Planck equation. Let us notice that
starting from an initial distribution which is a Dirac, i.e., choosing
$\Pini{x}=\dirac{x-\xzero}$, \cref{eq:Pfw} implies that
$P(x,t)=\Pfw{x,t|\xzero,\tzero}$. In other words, the transition
probability $\Pfw{x,t|\xzero,\tzero}$ is also the probability of
finding the process at $(x,t)$ when it started at $(\xzero,\tzero)$
with a Dirac distribution. For linear parabolic differential
equations, such as \cref{eq:FKfw}, $\Pfw{x,t|\xzero,\tzero}$ is also
the Green's function of the partial differential operator, here the
so-called adjoint Fokker-Planck operator~\cite{1979gsfb.book.....S}.

The transition probability of any Markov process also satisfies the
Chapman-Kolmogorov equation
\begin{equation}
\Pfw{x,t|y,s} = \int \Pfw{x,t|z,u} \Pfw{z,u|y,s} \dd{z},
\label{eq:CK}
\end{equation}
with $t > u > s$. Differentiating with respect to $u$ gives
\begin{equation}
\int \pdv{u}\left[\Pfw{x,t|z,u}\right]  \Pfw{z,u|y,s}\dd{z} = - \int
\Pfw{x,t|z,u} \pdv{u} \Pfw{z,u|y,s} \dd{z},
\label{eq:CKint}
\end{equation}
where the partial derivative on the right-hand side is given by
\cref{eq:FKfw}. Plugging
\cref{eq:FKfw} into \cref{eq:CKint}, and performing integrations by
parts, up to some vanishing boundary terms, one obtains the so-called
``backward'' Kolmogorov equation
\begin{equation}
-\pdv{s}\Pfw{x,t|y,s} =  F(y,s) \pdv{y}\Pfw{x,t|y,s} + \dfrac{1}{2}
G^2(y,s) \pdv[2]{y} \Pfw{x,t|y,s},
\label{eq:FKbw}
\end{equation}
with $t > s$. This is a partial differential equation for the same
transition probability distribution $\Pfw{x,t|y,s}$ as in
\cref{eq:FKfw} but, here, the time variable over which we
differentiate is the conditional time $s < t$. The backward equation
may be used to assess how likely is to be in a given state $(x,t)$
when one allows the previous states $(y,s)$ to vary. However,
\emph{both} \cref{eq:FKfw,eq:FKbw} describe a stochastic process
moving forward in times as $t>s$.

\subsection{Reverse Fokker-Planck equation}

Let us now consider the joint probability distribution
$\Pjt{x,t;y,s|\xzero,\tzero}$ to obtain the states $(x,t)$ and $(y,s)$
starting from a Dirac distribution at $(\xzero,\tzero)$. Using the
product rule, it reads
\begin{equation}
\Pjt{x,t;y,s|\xzero,\tzero} = \Pfw{y,s|\xzero,\tzero} \Pfw{x,t|y,s;\xzero,\tzero} ,
\label{eq:Pjtdef}
\end{equation}
with $t>s>\tzero$. Differentiating \cref{eq:Pjtdef} with respect to $s$,
using \cref{eq:FKfw,eq:FKbw}, one gets, after some algebra, a
Fokker-Planck equation for the joint
probability~\cite{nagasawa1993schroedinger,Anderson1982}
\begin{equation}
  \begin{aligned}
-\pdv{s} \Pjt{x,t;y,s|\xzero,\tzero} & = \pdv{y} \Bigg\{\left[F(y,s) - 2 G(y,s)
  \pdv{G(y,s)}{y} - G^2(y,s) \pdv{\ln \Pfw{y,s|\xzero,\tzero}}{y}
  \right] \\ & \times \Pjt{x,t;y,s|\xzero,\tzero} \Bigg\} + \dfrac{1}{2} \pdv[2]{y}\left[G^2(y,s)
  \Pjt{x,t;y,s|\xzero,\tzero}\right].
  \end{aligned}
\label{eq:FKjt}
\end{equation}
Using again the product rule, one can rewrite the joint probability
distribution as
\begin{equation}
\Pjt{x,t;y,s|\xzero,\tzero} = \Pof{y,s|x,t;\xzero,\tzero} \Pfw{x,t|\xzero,\tzero},
\label{eq:Pjtrule}
\end{equation}
still with $t>s>\tzero$. The first term in the right-hand side is
referring to a state $(y,s)$ conditioned by a future state $(x,t)$,
having started from $(\xzero,\tzero)$. It is precisely the transition
probability distribution of the time-reversed stochastic process
associated with \cref{eq:itodiff}~\cite{Nagasawa1964}. In order to
avoid any confusion, let us use the notation
\begin{equation}
\Prv{y,s|x,t;\xzero,\tzero} \equiv
\dfrac{\Pjt{x,t;y,s|\xzero,\tzero}}{\Pfw{x,t|\xzero,\tzero}} =
\dfrac{\Pfw{y,s|\xzero,\tzero}
  \Pfw{x,t|y,s;\xzero,\tzero}}{\Pfw{x,t|\xzero,\tzero}}\,,
\label{Prvdef}
\end{equation}
where, most importantly, $s<t$. From \cref{eq:FKjt}, one can
immediately derive the ``reverse'' Fokker-Planck equation\footnote{Not
to be confused with the backward Kolmogorov equation in \cref{eq:FKbw}.}
\begin{equation}
  \begin{aligned}
-\pdv{s} \Prv{y,s|x,t;\xzero,\tzero} & = \pdv{y} \Bigg\{\left[F(y,s) -
  2 G(y,s) \pdv{G(y,s)}{y} - G^2(y,s) \pdv{\ln
    \Pfw{y,s|\xzero,\tzero}}{y} \right] \\ & \times
\Prv{y,s|x,t;\xzero,\tzero} \Bigg\} + \dfrac{1}{2}
\pdv[2]{y}\left[G^2(y,s) \Prv{y,s|x,t;\xzero,\tzero}\right].
  \end{aligned}
\label{eq:FKrv}
\end{equation}
A few remarks are in order. Firstly, the reverse Fokker-Planck
equation depends explicitly on the forward transition probability
distribution $\Pfw{y,s|\xzero,\tzero}$. As a result, the determination
of $\Prv{y,s|x,t;\xzero,\tzero}$ requires to first solve
\cref{eq:FKfw}. Secondly, the reverse process remains always
conditioned by the initial state $(\xzero,\tzero)$ of the forward
process. This is not surprising as a time inversion requires the
reverse process to end at the state $(\xzero,\tzero)$. Finally,
\cref{eq:FKrv} being a Fokker-Planck equation describing a Markov
process, it stems from an It\^o stochastic differential
equation~\cite{Chung1969ToRA,Haussmann1985,Pardoux1985,Haussmann1986}
\begin{equation}
\dd{\Xrv} = \Frv\left[\Xrv(s),s\right]\dd{s} + \Grv\left[\Xrv(s),s\right] \dd{W(s)},
\label{eq:xrv}
\end{equation}
where $\Grv = G$ and
\begin{equation}
\Frv(y,s) = F(y,s) - \dfrac{1}{\Pfw{y,s|\xzero,\tzero}}
\pdv{\left[G^2(y,s) \Pfw{y,s|\xzero,\tzero}\right]}{y}.
\label{eq:frv}
\end{equation}
Given $\Pfw{y,s|\xzero,\tzero}$, \cref{eq:xrv,eq:frv} can be used to
numerically draw realisations of the time-reversed process.
From \cref{eq:FKrv}, it appears that the state $(x,t)$ now acts as a
conditioner, $t$ being the time from which the process is
reversed. One of the simplest time one can choose for $t$ is the
lifetime of the process, a random time, and this ensures that the
reverse process preserves the strong Markov
property~\cite{chung2005markov}. It is also possible to reverse from a
fixed time, or various other random times belonging to a class of
so-called co-optional times~\cite{Nagasawa1964}.

In a forward picture, the time-reversal could be viewed as performing
a path-by-path selection of realisations that are conditioned to meet
a certain value $x$ after having evolved for a certain time
$t$. Therefore, it should come with no surprise that it is closely
related to conditioned diffusion processes~\cite{Baudoin2002,
  Orland2011, Majumdar_2015, Ardorisio2018}. Mathematically, the
conditioning which emerges by enforcing some specific constraints is
performed through the so-called $h$-transforms~\cite{Doob1957,
  doob2001classical, Rogers_Williams_2000}, and they lead to
Fokker-Planck equations with modified drift, in a similar spirit to
\cref{eq:FKrv}. A formalism of constrained stochastic inflation has
been presented in Ref.~\cite{Tokeshi:2023swe}, and applied to
multifield scenarios to show that stochastic inflation converges to be
single-field dominated. Although time-reversing a stochastic process
is formally different than conditioning, the way to implement it is,
in practice, closely linked to $h$-transforms techniques. An in-depth
analysis of $h$-transforms and their connections with time reversal
and last passage times can be found in Ref.~\cite{chung2005markov}.

\section{Semi-infinite quantum diffusion domain}
\label{sec:sols}

Let us now apply the techniques summarized in the previous section to
stochastic inflation in the semi-infinite quantum diffusion domain. As
already mentioned, we approximate the field potential at $\phi>\phiqw$
by a constant and \cref{eq:langevin} simplifies to
\begin{equation}
\dd{\phi} = \dfrac{H}{2\pi} \xi \dd{N}.
\label{eq:itoflat}
\end{equation}
The first Friedmann-Lema\^itre equation gives $H^2=\Hinf^2 \simeq V/3$
which is therefore constant (the quantum diffusion on the velocity
being negligible). The stochastic process described by
\cref{eq:itoflat} is the simplest of all, a pure Brownian motion with
\begin{equation}
G(\phi,N) = \Ginf \equiv \dfrac{\Hinf}{2\pi}\,,
\end{equation}
with, however, an absorbing boundary located at the quantum wall
$\phi=\phiqw$.

\subsection{Forward stochastic inflation}

Denoting the forward transition probability by
$\Pfw{\phi,N|\phizero,\Nzero}$, it is solution of \cref{eq:FKfw} with
vanishing drift and constant diffusion, i.e.,
\begin{equation}
  \pdv{\Pfw{\phi,N|\phizero,\Nzero}}{N} = \dfrac{1}{2} \Ginf^2
  \pdv[2]{\Pfw{\phi,N|\phizero,\Nzero}}{\phi}\,.
\label{eq:stofw}
\end{equation}
Enforcing both the boundary and initial conditions
\begin{equation}
  \Pfw{\phiqw,N|\phizero,\Nzero}=0\,, \qquad
  \Pfw{\phi,\Nzero|\phizero,\Nzero} = \dirac{\phi-\phizero},
\end{equation}
the solution to \cref{eq:stofw} can be straightforwardly obtained using
Fourier transform and the method of images. One gets
\begin{equation}
\Pfw{\phi,N|\phizero,\Nzero} = \dfrac{1}{\sqrt{2\pi} \Ginf
  \sqrt{N-\Nzero}} \left[e^{-\frac{\left(\phi-\phizero\right)^2}{2
        \Ginf^2 \left(N-\Nzero\right)}} - e^{-\frac{\left(\phi -2
      \phiqw + \phizero\right)^2}{2
        \Ginf^2 \left(N-\Nzero\right)}} \right].
\label{eq:Pfwsto}
\end{equation}
We wish to time-reverse the coarse-grained field evolution with
respect to $\Nqw$, the first passage time of each realisation through
the quantum wall at $\phi=\phiqw$. Following Ref.~\cite{Ando:2020fjm},
the survival probability for the field to be within the quantum region
at {\efold} $N$ is given by
\begin{equation}
S(N|\phizero,\Nzero) =
\int_{\phiqw}^{+\infty}\Pfw{\phi,N|\phizero,\Nzero}\dd{\phi} = 1 -
\int_{\Nzero}^{N}\Plt{\Nqw|\phizero,\Nzero} \dd{\Nqw},
\label{eq:survival}
\end{equation}
where $\Plt{\Nqw|\phizero,\Nzero}$ gives the probability distribution
of $\Nqw$, the first passage time at $\phi=\phiqw$. Differentiating
\cref{eq:survival} with respect to $N$, one gets
\begin{equation}
\Plt{N|\phizero,\Nzero} = -\pdv{S(N|\phizero,\Nzero)}{N} =
\dfrac{G^2}{2}
\left[\eval{\pdv{\Pfw{\phi,N|\phizero,\Nzero}}{\phi}}_{\phiqw} -
  \eval{\pdv{\Pfw{\phi,N|\phizero,\Nzero}}{\phi}}_{+\infty} \right],
\end{equation}
where use has been made of \cref{eq:stofw}. Plugging \cref{eq:Pfwsto}
in this expression gives
\begin{equation}
\Plt{\Nqw|\phizero,\Nzero} = \dfrac{\phizero - \phiqw}{\sqrt{2\pi} G
  \left(\Nqw-\Nzero\right)^{3/2}}\,
e^{-\frac{\left(\phizero-\phiqw\right)^2}{2 G^2 \left(\Nqw-\Nzero\right)}}\,,
\label{eq:Pltsto}
\end{equation}
which is also the probability distribution of the lifetimes
$\Nqw-\Nzero$.

\subsection{Time-reversed stochastic inflation}
\label{sec:revsto}

One can time reverse stochastic inflation from $\Nqw$, the {\efold} at
which the field exits the quantum region, for each of its
realisations. This is a random variable having the probability
distribution of \cref{eq:Pltsto}. Let us define the time-reversed number
of {\efolds}
\begin{equation}
\rv{N} \equiv \Nqw - N,
\end{equation}
ranging from $\rv{N}=0$, when the field hits the quantum wall, to a
maximal value $\rv{\Nzero}=\Nqw-\Nzero$ which is the lifetime of the
process.  Forgetting about the forward picture, time-reversed
stochastic inflation describes the evolution of a stochastic field
$\phi$ emerging at $\phi=\phiqw$, at the time $\rv{N}=0$, and
randomly evolving towards a sink located at $\phi=\phizero$, which
is necessarily reached at the time $\rv{N}=\rv{\Nzero}$. This process is
completely characterised by the probability distribution function
$\Prv{\phi,\rv{N}|\phizero,\rv{\Nzero}}$, solution of
\cref{eq:FKrv}, i.e.,
\begin{equation}
  \begin{aligned}
\pdv{\Prv{\phi,\rv{N}|\phizero,\rv{\Nzero}}}{\rv{N}} & = -G^2 \pdv{\phi}
\left[ \pdv{\ln \Pfw{\phi,N|\phizero,\Nzero}}{\phi}
  \Prv{\phi,\rv{N}|\phizero,\rv{\Nzero}} \right] \\ & + \dfrac{1}{2}
G^2 \pdv[2]{\Prv{\phi,\rv{N}|\phizero,\rv{\Nzero}}}{\phi}\,.
\end{aligned}
\end{equation}
Using the explicit expression of $\Pfw{\phi,N|\phizero,\Nzero}$ given
by \cref{eq:Pfwsto}, this equation simplifies to the one of a Brownian
motion in presence of drift:
\begin{equation}
\pdv{\Prv{\phi,\rv{N}|\phizero,\rv{\Nzero}}}{\rv{N}} =
-\pdv{\phi}\left[\Frv(\phi,\rv{N})
  \Prv{\phi,\rv{N}|\phizero,\rv{\Nzero}}\right] + \dfrac{1}{2} G^2 \pdv[2]{\Prv{\phi,\rv{N}|\phizero,\rv{\Nzero}}}{\phi}\,.
\label{eq:FKrvsto}
\end{equation}
The friction term reads
\begin{equation}
\Frv(\phi,\rv{N}) =
-\dfrac{\rv{\phi}-\rv{\phizero}}{\rv{\Nzero}-\rv{N}} +
\dfrac{2\rv{\phizero}}{{\rv{\Nzero}-\rv{N}}}
  \dfrac{e^{-\frac{2 \rv{\phizero} \rv{\phi}}{G^2 \left(\rv{\Nzero}-\rv{N}\right)}}}{1 -
    e^{-\frac{2 \rv{\phizero} \rv{\phi}}{G^2\left(\rv{\Nzero} -\rv{N}\right)}}} \equiv
  \Frvzero + \Frvinfty \,,
\label{eq:Frvsto}
\end{equation}
where we have introduced field values in reference to the
quantum wall, i.e.,
\begin{equation}
\rv{\phi} \equiv \phi - \phiqw.
\end{equation}
The form of the friction term can be qualitatively understood by
remarking that, for $\rv{\phi} \to 0$, i.e., close to the quantum
wall, one has
\begin{equation}
\Frv \to \Frvinfty \simeq \dfrac{G^2}{\rv{\phi}}\,.
\label{eq:Frvinfty}
\end{equation}
This represents an infinite repelling force from the quantum wall
ensuring, as it should, that the (reverse) random process once entered
into the domain $\phi>\phiqw$ cannot turn back and escape. Far from
the quantum wall, for $\rv{\phi}\rv{\phizero}/G^2 \gg
\rv{\Nzero}-\rv{N}$, one has
\begin{equation}
\Frv \to \Frvzero = - \dfrac{\rv{\phi} -\rv{\phizero}}{\rv{\Nzero}-\rv{N}}\,.
\label{eq:Frvzero}
\end{equation}
Such a friction term describes an elastic force in field space, the
spring begin attached at $\phizero$, with an elasticity coefficient
growing with time as $1/(\rv{\Nzero}-\rv{N})$, i.e., becoming
infinitely attractive when the lifetime is reached. Clearly, this term
ensures that the reverse process ends at $\phi=\phizero$ for $\rv{N} =
\rv{\Nzero}$. 

\begin{figure}
\begin{center}
  \includegraphics[width=\onefigw]{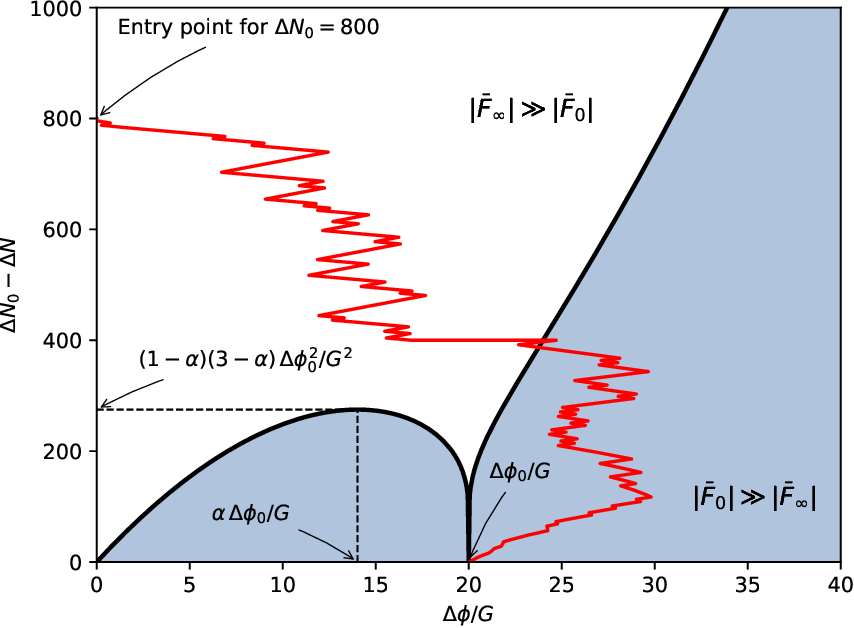}
\caption{Sketch of a time-reversed stochastic inflation realisation
  when $\rv{\Nzero} \gg \rv{\phizero^2}/G^2$. The field enters the
  quantum domain for $\rv{\phi}=0$ ($\phi=\phiqw$), at the time
  $\rv{N}=0$. In the white region, the repulsive drift $\Frvinfty$
  pushes the field away from $\rv{\phi}=0$ whereas, in the blue
  domains, an elastic attractive force $\Frvzero$ captures the field
  and drives it to $\rv{\phizero}$. The sink at $\rv{\phizero}$ is
  necessarily reached at $\rv{N}=\rv{\Nzero}$, when the number of
  time-reversed {\efolds} has exhausted the lifetime of the forward
  process. In the opposite limit, $\rv{\Nzero} \ll
  \rv{\phizero^2}/G^2$, the process is mostly driven by the attractive
  elastic force towards $\rv{\phizero}$, white regions are almost
  non-existent (see also \cref{fig:rvprob}).}
\label{fig:rvdomains}
\end{center}
\end{figure}

In \cref{fig:rvdomains}, we have represented, in a field-time plot,
the two basins of repulsion and attraction, together with a sketch of
one of the realisation of the time-reverse process. The solid black
line is the separatrix between the two regions, the solution of
$\abs{\Frvzero}=\abs{\Frvinfty}$, i.e.,
\begin{equation}
\rv{\Nzero}-\rv{N} = \dfrac{2 \rv{\phizero}
    \rv{\phi}}{G^2 \ln\left(1+\dfrac{2\rv{\phizero}}{\abs{\rv{\phi}-\rv{\phizero}}}\right)}\,.
\end{equation}
This curve has a local maximum in the region $\rv{\phi}<\rv{\phizero}$, located at
\begin{equation}
\rv{\phimax} = \alpha \, \rv{\phizero}, \qquad \rv{\Nzero}-\rv{\Nmax} =
\left(1-\alpha\right)\left(3 - \alpha \right) \dfrac{\rv{\phizero}^2}{G^2}\,,
\end{equation}
where $\alpha \simeq 0.701$ is a constant that has to be determined
numerically. As a result, for $\rv{\Nzero} \ll \rv{\phizero}^2/G^2$,
most of the field-time domain is dominated by the friction term
$\Frvzero$ and, up to some narrow band close to the quantum wall, the
field is strongly driven towards $\rv{\phizero}$. This is expected. In
the forward picture, the process is Brownian, and conditioning by a
lifetime smaller than the typical diffusion time $\rv{\phizero}^2/G^2$
implies that we select the fastest trajectories jumping from
$\phizero$ to $\phiqw$. On the contrary, for $\rv{\Nzero} \gg
\rv{\phizero}^2/G^2$, the white region is present and the field may
freely diffuse before getting trapped by $\rv{\phizero}$.

\subsection{Exact solution}
\label{sec:exactsol}

It is possible to find approximate solutions of \cref{eq:FKrvsto},
using either \cref{eq:Frvinfty}, \cref{eq:Frvzero}, or both.  However,
an exact solution of \cref{eq:FKrvsto}, with a friction term of the
form \cref{eq:Frvsto}, has been recently found in
Ref.~\cite{Mazzolo2024}. Here, we simply quote the result and detail
the astonishing derivation of this solution from the Girsanov's
theorem in \cref{sec:girsanov}.  The solution of \cref{eq:FKrvsto},
satisfying the initial and boundary conditions
\begin{equation}
\Prv{\phi,\rv{N}=0|\phizero,\rv{\Nzero}} = \dirac{\rv{\phi}},\quad
\Prv{\phi,\rv{N}=\rv{\Nzero}|\phizero,\rv{\Nzero}} = \dirac{\rv{\phi}-\rv{\phizero}},
\label{eq:Prvibc}
\end{equation}
reads
\begin{equation}
  \begin{aligned}
\Prv{\phi,\rv{N}|\phizero,\rv{\Nzero}} & = \sqrt{\dfrac{2}{\pi}}
\dfrac{\rv{\Nzero}^{3/2}}{\rv{N}^{3/2} G \sqrt{\rv{\Nzero} - \rv{N}}}
\dfrac{\rv{\phi}}{\rv{\phizero}}
\sinh\left[\dfrac{\rv{\phizero}\rv{\phi}}{G^2\left(\rv{\Nzero} -
    \rv{N} \right)}\right] \\ & \times \exp\left[-\dfrac{\rv{\Nzero}^2
    \rv{\phi}^2 + \rv{N}^2 \rv{\phizero}^2}{2 G^2 \rv{N}\rv{\Nzero}
    \left(\rv{\Nzero}-\rv{N}\right)}\right].
  \end{aligned}
\label{eq:solrv}
\end{equation}
We can make this expression more readable by measuring field
excursion values in unit of the quantum diffusion strength $G$ and counting
reverse {\efolds} numbers in unit of the lifetime, namely
\begin{equation}
\chi \equiv \dfrac{\rv{\phi}}{G}\,, \qquad \tau \equiv
\dfrac{\rv{N}}{\rv{\Nzero}}\,.
\end{equation}
One gets
\begin{equation}
\Prv{\phi,\rv{N}|\phizero,\rv{\Nzero}}  =
\dfrac{\sqrt{2/\pi}}{\tau^{3/2} \sqrt{1-\tau}} \,
\dfrac{\chi}{\chizero G \sqrt{\rv{\Nzero}}}
\sinh\left[\dfrac{\chi\chizero}{(1-\tau)\rv{\Nzero}}\right]
e^{-\frac{\chi^2 + \tau^2 \chizero^2}{2 \tau(1-\tau)\rv{\Nzero}}},
\label{eq:solsimple}
\end{equation}
with $0<\tau<1$.

\begin{figure}
\begin{center}
  \includegraphics*[width=\onefigw]{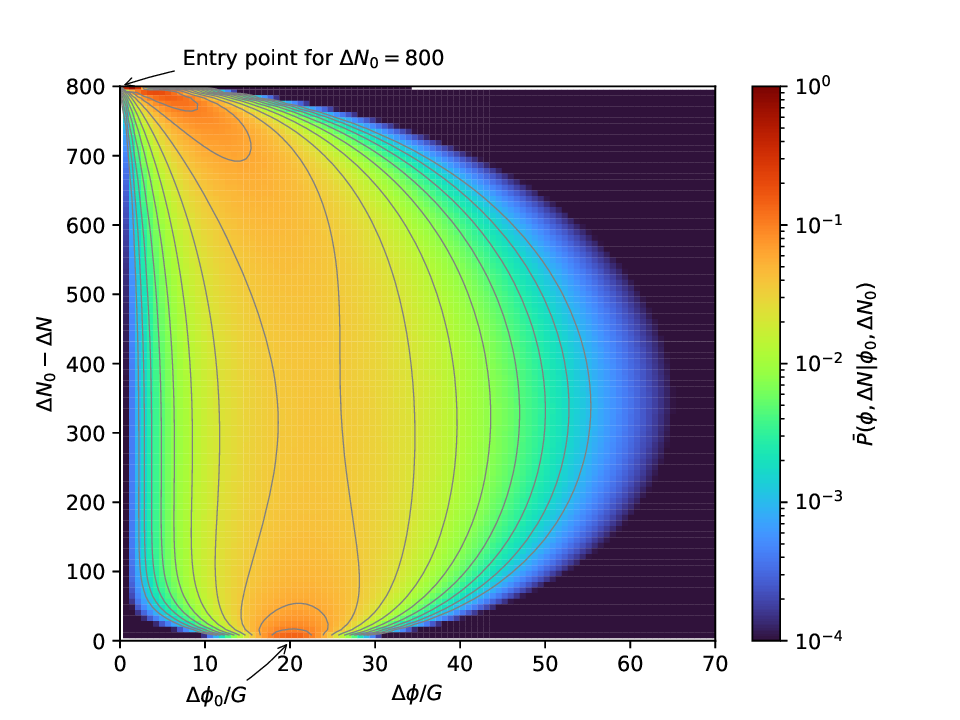}
  \includegraphics*[width=\onefigw]{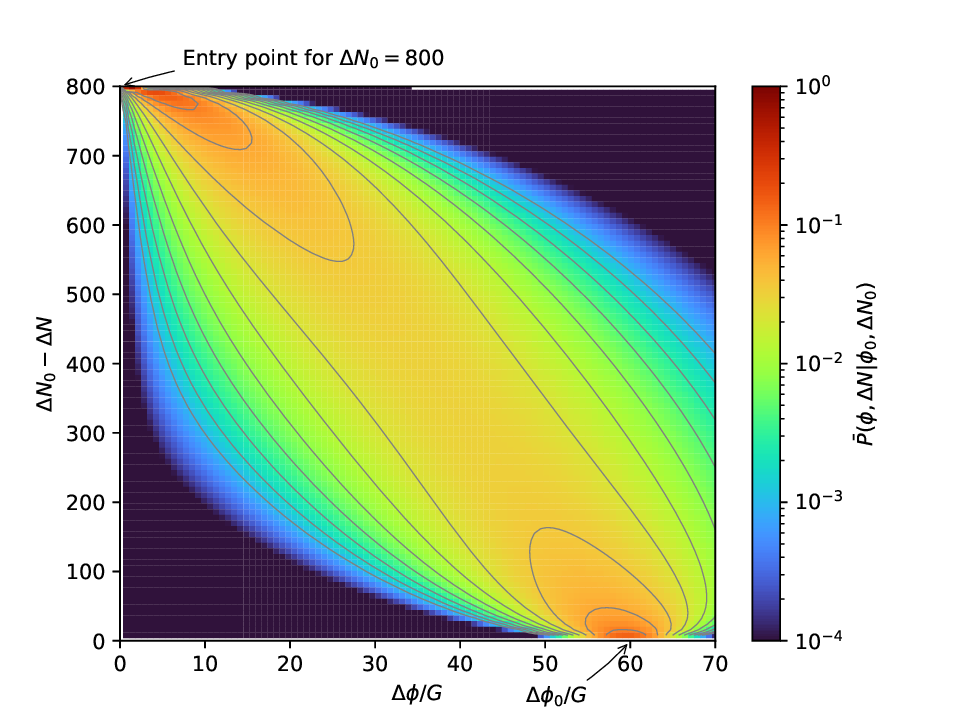}
\caption{Contour plots of the reverse probability distribution
  $\Prv{\phi,\rv{N}|\phizero,\rv{\Nzero}}$, for an arbitrary lifetime
  set at $\rv{\Nzero}=800$ {\efolds}, with $\rv{\phizero}=20G$ (upper
  panel) and $\rv{\phizero}=60G$ (lower panel). For $\rv{\phizero}\gg G
  \sqrt{\rv{\Nzero}}$, diffusion is reduced and the field flows from
  the quantum wall to the sink.}
\label{fig:rvprob}
\end{center}
\end{figure}

Contour plots of \cref{eq:solsimple} in the field-time plane are
represented in \cref{fig:rvprob}, in the two regimes $\rv{\Nzero} >
\chizero^2$ (upper panel) and $\rv{\Nzero} < \chizero^2$ (lower
panel). The narrower contours in the latter case confirms the
qualitative discussion of \cref{sec:revsto}, the field flows from
$\rv{\phi}=0$ to $\rv{\phizero}$ in shorter times. For the limit
$\rv{\Nzero} \gg \chizero^2$ and $\tau \ll 1$, \cref{eq:solsimple}
readily simplifies into
\begin{equation}
\Prv{\phi,\rv{N}|\phizero,\rv{\Nzero}} \simeq \sqrt{\dfrac{2}{\pi}}
\dfrac{\chi^2}{G \left(\tau\rv{\Nzero}\right)^{3/2}}
e^{-\dfrac{\chi^2}{2 \tau\rv{\Nzero}}}.
\end{equation}
This is the probability distribution of a Brownian motion in presence
of a repelling friction in $G^2/\rv{\phi}$. As expected, the
probability of finding the field close to the quantum wall is indeed
vanishing, as visible in \cref{fig:rvprob}.

In the next section, we use the exact solution of \cref{eq:solsimple}
to extract the generated curvature fluctuations using an adapted
version of the stochastic $\deltaN$-formalism.

\section{Reverse-time stochastic $\deltaN$-formalism}
\label{sec:rvdeltaN}

The $\deltaN$-formalism relates the distribution of the curvature
fluctuations $\zeta$, on super-Hubble scales, to the fluctuations in
the amount of expansion by $\zeta = N(\bx) - \bar{N}$ where $\bar{N}$
is the unperturbed number of {\efolds}~\cite{Sasaki:1995aw,
  Sasaki:1998ug, Wands:2000dp, Lyth:2004gb, Lyth:2005fi}. In the context of
stochastic inflation, $\zeta$ and $N$ are being promoted to stochastic
variables and the previous relation is replaced
by~\cite{Enqvist:2008kt, Fujita:2013cna,Fujita:2014tja, Vennin:2015hra, Pattison:2017mbe}
\begin{equation}
    \zeta = N-\Nzero - \ev{N-\Nzero}.
\label{eq:zetafw}
\end{equation}
In the forward stochastic inflation formalism, difficulties may arise
in the determination of $\ev{N-\Nzero}$. In particular, in the case we are
interested in, the semi-infinite flat quantum domain, \cref{eq:Pltsto} is
a Levy distribution and $\ev{\Nqw-\Nzero}$ is infinite.

For time-reversed stochastic inflation, the random variables are, a
priori, $\rv{N}=\Nqw-N$ and $\rv{\Nzero}=\Nqw-\Nzero$ such that, from
\cref{eq:zetafw}, one gets $\zeta = \rv{\Nzero} - \rv{N} + \ev{\rv{N}}
- \ev{\rv{\Nzero}}$.  However, the expectation value for the
time-reversed processes has not the same meaning as the one of
\cref{eq:zetafw}. It is now in reference to realisations conditioned
by the lifetime $\rv{\Nzero}$, as can be seen in the very definition
of $\Prv{\phi,\rv{N},|\phizero,\rv{\Nzero}}$. Thus,
$\ev{\rv{\Nzero}}=\rv{\Nzero}$ and the time-reverse curvature
fluctuations are simply given by
\begin{equation}
\zeta = \ev{\rv{N}} - \rv{N}.
\label{eq:zetarv}
\end{equation}
This form emphasizes that the reverse-time stochastic
$\deltaN$-formalism automatically gives the statistics of the
curvature perturbation as measured by an observer lying on the
end-of-inflation hypersurface. Moreover, the conditioning by the
lifetime $\rv{\Nzero}$ ensures that the mean values are not becoming
infinite due to the possible eternally inflating realisations. As
discussed in the introduction, such a conditioning, by $\rv{\Nzero}$,
can be viewed as a partitioning of all the forward processes by the
lifetimes and \cref{eq:zetarv} gives the curvature fluctuations
generated within each partition. Of course, one has to add the
contribution of all the partitions to get $\zeta$ over the whole set of
stochastic processes, i.e., we will ultimately marginalise
\cref{eq:zetarv} over the lifetimes.

Let us now detail how to derive the probability distribution for the
curvature fluctuations generated in the semi-infinite quantum
diffusion region. From \cref{eq:zetarv}, at given lifetime, the
time-reverse approach gives us the joint probability distribution
\begin{equation}
\Pof{\phi,\zeta|\phizero,\rv{\Nzero}} =
\Pof{\phi,\rv{N}=\ev{\rv{N}}-\zeta|\phizero,\rv{\Nzero}},
\label{eq:PofPhiAndZetadef}
\end{equation}
where $\rv{N}$ now refers to a stochastic variable and one needs to
evaluate $\ev{\rv{N}}$. In the reverse Fokker-Planck equation, the
time variable is a parameter and one may wonder how $\rv{N}$ is
suddenly promoted to be a stochastic quantity even though the lifetime
is fixed. This can be understood by considering a slice of constant
field value in \cref{fig:rvprob}. A given $\rv{\phi}$ can be reached
by many random trajectories, at different number of {\efolds}
$\rv{N}$. As such, $\rv{N}$ can be viewed as a stochastic variable,
having a probability distribution depending on $\rv{\phi}$. According
to the stochastic $\deltaN$-formalism, no curvature perturbation is
generated without randomness in $\rv{N}$. Indeed, if \cref{fig:rvprob}
were showing a single classical trajectory between the quantum wall
and $\rv{\phizero}$, the distribution of $\rv{N}$ would be a Dirac
distribution centred at the classical value, say $\rv{N}(\rv{\phi})$,
for all $\rv{\phi}$, implying $\ev{\rv{N}}=\rv{N}$ and $\zeta$ would
vanish. By this argument, one sees the relevance of measuring the
fluctuations of $\rv{N}$ in reference to $\ev{\rv{N}}$.

In the next section, we first determine the mean number of {\efold}
$\ev{\rv{N}}$, at given field value and lifetime. Then, in
\cref{sec:PzetaGivenLT}, we marginalise \cref{eq:PofPhiAndZetadef}
over the field values $\phi$ to obtain the distribution of $\zeta$ at
given lifetime, i.e., within each partition. Finally, in
\cref{sec:pzeta}, we marginalise over all the possible lifetimes to
obtain the final quantum-generated curvature fluctuations.

\subsection{Mean number of e-folds}
\label{sec:taumean}
In order to determine $\ev{\rv{N}}$, we first need to calculate the
probability distribution
\begin{equation}
\Pof{\rv{N}|\phi,\phizero,\rv{\Nzero}} =
\dfrac{\Prv{\phi,\rv{N}|\phizero,\rv{\Nzero}}}{\Pof{\phi|\phizero,\rv{\Nzero}}}\,,
\label{eq:PofNgivenphi}
\end{equation}
where the denominator is a normalisation factor given by
\begin{equation}
\Pof{\phi|\phizero,\rv{\Nzero}} = \int_0^{\rv{\Nzero}}
\Prv{\phi,\rv{N}|\phizero,\rv{\Nzero}} \dd{\rv{N}} = \rv{\Nzero}
\int_0^1 \Prv{\phi,\tau \rv{\Nzero}|\phizero,\rv{\Nzero}} \dd{\tau}.
\label{eq:Pofphidef}
\end{equation}
The only assumption used in \cref{eq:PofNgivenphi} is that
$\Pof{\phi,\rv{N}|\phizero,\rv{\Nzero}} \propto
\Prv{\phi,\rv{N}|\phizero,\rv{\Nzero}}$, which may be considered
as inherent to the reverse-time stochastic $\deltaN$-formalism. The mean
value $\ev{\rv{N}}$ is then given by
\begin{equation}
\eval{\ev{\rv{N}}}_{\rv{\phi},\phizero,\rv{\Nzero}} = \int_0^{\rv{\Nzero}} \rv{N}\,
\Pof{\rv{N}|\phi,\phizero,\rv{\Nzero}} \dd{\rv{N}},
\label{eq:meanNdef}
\end{equation}
which is a functional of $\rv{\phi}$, $\rv{\phizero}$, $\rv{\Nzero}$.

Let us first determine the normalisation factor of
\cref{eq:Pofphidef}. In terms of $\tau$ and $\chi$, expanding the
hyperbolic sine in \cref{eq:solsimple}, one gets
\begin{equation}
\Prv{\phi,\tau\rv{\Nzero}|\phizero,\rv{\Nzero}} =
\dfrac{\chi}{\chizero \sqrt{2\pi}G \sqrt{\rv\Nzero}}
\dfrac{1}{\tau^{3/2}\sqrt{1-\tau}} \left[ e^{-\frac{\left(\chi-\tau
      \chizero\right)^2}{2 \rv{\Nzero} \tau(1-\tau)}} -
  e^{-\frac{\left(\chi + \tau \chizero\right)^2}{2 \rv{\Nzero}
      \tau(1-\tau)}} \right],
\label{eq:PrvTwoGaussians}
\end{equation}
which is the difference of two Gaussians and has to be integrated over
$\tau$. With an adequate change of variable, $z=1/\tau -1$, both terms
are known integrals of the form~\cite{gradshteyn2007}
\begin{equation}
\int_0^{+\infty} \dfrac{e^{-\frac{\beta}{z} - \gamma z}}{\sqrt{z}} \dd{z} =
\sqrt{\dfrac{\pi}{\gamma}}\, e^{-2 \sqrt{\beta \gamma}},
\label{eq:GR_p3.475eq15}
\end{equation}
where $\beta$ and $\gamma$ are any positive constant. After some algebra, one
gets
\begin{equation}
\Pof{\phi|\phizero,\rv{\Nzero}} = \dfrac{\rv{\Nzero}}{\chizero G} \left[
e^{-\frac{\chi}{\rv{\Nzero}}\left(\left|\chi-\chizero\right| +
    \chi-\chizero\right)} -
e^{-\frac{2\chi}{\rv{\Nzero}}\left(\chi+\chizero\right)} \right].
\label{eq:Pofphi}
\end{equation}

\begin{figure}
\begin{center}
  \includegraphics[width=\onefigw]{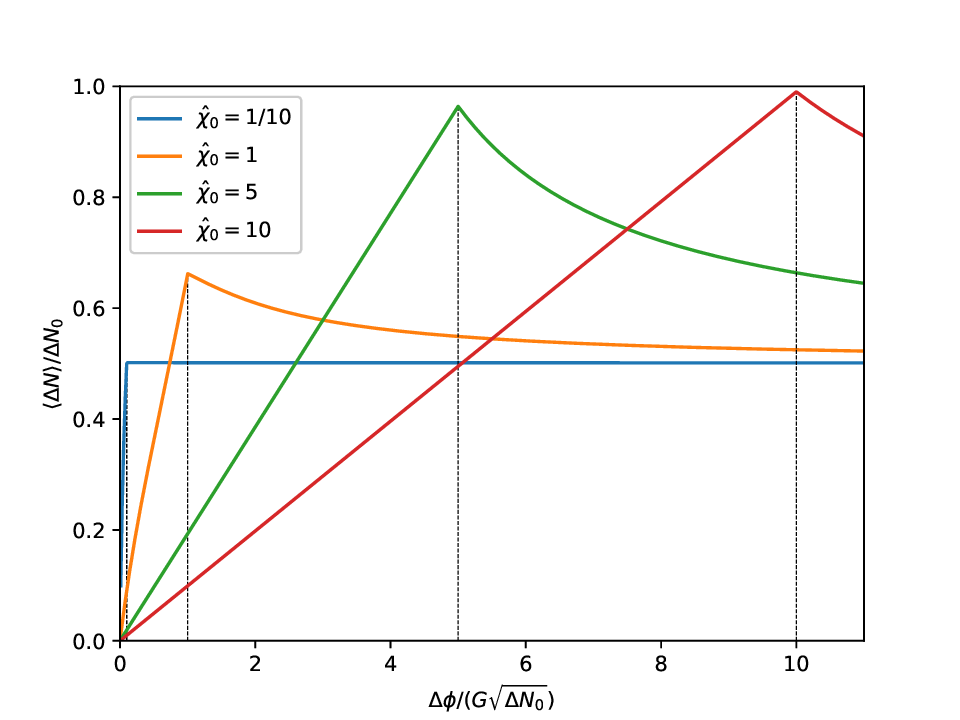}
\caption{Mean number of {\efolds} $\ev{\tau}=\ev{\rv{N}}/\rv{\Nzero}$,
  in unit of the lifetime, as a function of the field $\chihat =
  \rv{\phi}/(G\sqrt{\rv{\Nzero}})$, plotted for various values of
  $\chizerohat = \rv{\phizero}/(G\sqrt{\rv{\Nzero}})$.  Generically,
  $\ev{\tau}$ increases for $\rv{\phi}<\rv{\phizero}$ and decreases to
  asymptote $1/2$ for $\rv{\phi} > \rv{\phizero}$. However, for small
  $\chizerohat$, i.e. $\phizero$ small and/or very large lifetimes
  $\rv{\Nzero}$, one has $\ev{\tau}= 1/2$ in almost all the field
  domain.}
\label{fig:taumean}
\end{center}
\end{figure}

The mean value $\ev{\rv{N}}$, expressed in unit of the lifetime, reads
\begin{equation}
\ev{\tau} = \dfrac{\ev{\rv{N}}}{\rv{\Nzero}} = \rv{\Nzero} \int_0^1
\tau \, \Pof{\tau\rv{\Nzero}|\phi,\phizero,\rv{\Nzero}}
\dd{\tau}.
\end{equation}
Because the normalisation factor, \cref{eq:Pofphi}, does not depend on $\tau$,
one has now to determine the integral of $\tau \times
\Prv{\phi,\tau\rv{\Nzero}|\phizero,\rv{\Nzero}}$. Formally, this
consists in replacing the term in $1/\tau^{3/2}$ by $1/\tau^{1/2}$ in
the integrand given by \cref{eq:PrvTwoGaussians}. Using the same
change of variable, $z=1/\tau -1$, one has to calculate integrals of
the form
\begin{equation}
K(\beta,\gamma) = \int_0^{+\infty} \dfrac{e^{-\frac{\beta}{z} - \gamma
    z}}{\sqrt{z} \left(1+z\right)} \dd{z},
\end{equation}
where $\beta>0$ and $\gamma>0$. This class of integrals is not
documented in standard textbooks, but, as shown in
\cref{sec:Kintegral}, they can be determined exactly by using infinite
series resummation and we find
\begin{equation}
K(\beta,\gamma) = \pi e^{\beta+\gamma} \erfc\left(\sqrt{\beta} + \sqrt{\gamma}\right).
\end{equation}
Here $\erfc(x)=1-\erf(x)$ is the complementary error function. After
some algebra, one finally gets
\begin{equation}
\ev{\tau} = \dfrac{\ev{\rv{N}}}{\rv{\Nzero}}= \sqrt{\dfrac{\pi}{2}} \,
  \chihat \, e^{\frac{\chizerohat^2}{2}} \, \dfrac{\erf\left(\dfrac{2
      \chihat + \chizerohat}{\sqrt{2}}\right) -
    \erf\left(\dfrac{\chihat +
      \left|\chihat-\chizerohat\right|}{\sqrt{2}}\right)}{e^{-\chihat\left(\left|\chihat-\chizerohat\right|
      + \chihat-\chizerohat\right)} -
    e^{-2\chihat\left(\chihat+\chizerohat\right)}}\,,
\label{eq:taumean}
\end{equation}
where use has been made of \cref{eq:Pofphi}. Notice that all explicit
dependencies in $\rv{\Nzero}$ have been absorbed into the rescaled
dimensionless field values
\begin{equation}
\chihat \equiv \dfrac{\chi}{\sqrt{\rv{\Nzero}}} =
\dfrac{\rv{\phi}}{G\sqrt{\rv{\Nzero}}}, \qquad \chizerohat \equiv  \dfrac{\chizero}{\sqrt{\rv{\Nzero}}} =
\dfrac{\rv{\phizero}}{G\sqrt{\rv{\Nzero}}}\,.
\label{eq:chihatdef}
\end{equation}
We have plotted $\ev{\tau}$ as a function of the field values $\chihat$ in
\cref{fig:taumean}. It is a monotonic increasing function of $\chihat$
in the domain $\rv{\phi}<\rv{\phizero}$ and a decreasing one, towards
the asymptotic value $1/2$ at $\rv{\phi}>\rv{\phizero}$. Such a
behaviour is consistent with the two-dimensional contours of the
probability distribution plotted in \cref{fig:rvprob}. Using the error
function expansion at large argument, the behaviour of $\ev{\tau}$ for
$\chizerohat \gg 1$, and for $\chihat>\chizerohat$, is given by
\begin{equation}
\ev{\tau} \simeq \dfrac{2\chihat^2}{4\chihat^2 - \chizerohat^2} +
\dfrac{\chihat\chizerohat}{\left(4\chihat^2 -\chizerohat^2\right)\tanh\left(2\chihat\chizerohat\right)}\,,
\end{equation}
which asymptotes $1/2$ for $\chihat \gg \chizerohat$. In the opposite
limit, $\chizerohat \ll 1$, with $\chihat<\chizerohat$, one  obtains
\begin{equation}
\ev{\tau} \simeq \dfrac{\chihat}{\chihat + \chizerohat}\,,
\end{equation}
which is nearly linear for $\chihat \ll \chizerohat$. The exact curves
plotted in \cref{fig:taumean} show that this approximation is actually
very good even when $\chizerohat$ is not small. Indeed, for
$\chizerohat \gg 1$, still with $\chihat < \chizerohat$, one can
Taylor expand the error functions in \cref{eq:taumean} around
$\chizerohat/\sqrt{2}$ and use their large argument expansion to show that
$\ev{\tau} \simeq \chihat/\chizerohat$. Finally, the limit
$\chizerohat \ll 1$ and $\chihat>\chizerohat$ can also be worked out
to show that $\ev{\tau} \to 1/2$, as visible in \cref{fig:taumean}.

\subsection{Curvature fluctuations at given lifetime}
\label{sec:PzetaGivenLT}

From the previous discussion, the joint probability distribution
$\Pof{\phi,\zeta|\phizero,\rv{\Nzero}}$ is given by
\cref{eq:solrv} upon replacing $\rv{N}$ by $\ev{\rv{N}}-\zeta$ where
$\ev{\rv{N}}$ is a function of $\rv{\phi}$, $\rv{\phizero}$ and
$\rv{\Nzero}$ given by \cref{eq:taumean}. In terms of the rescaled
quantities introduced earlier, the friendly version for a numerical
integration reads
\begin{equation}
  \begin{aligned}
\Pof{\phi,\zetahat|\phizero,\rv{\Nzero}} & =
\dfrac{1}{\sqrt{2\pi}G\sqrt{\rv{\Nzero}}} \dfrac{\chihat}{\chizerohat}
\dfrac{\heaviside{\ev{\tau}-\zetahat} - \heaviside{\ev{\tau} -
    \zetahat -1}}{\left(\ev{\tau}-\zetahat\right)^{3/2}
  \sqrt{1-\ev{\tau} + \zetahat}}  \\ & \times \exp\left\{-\dfrac{\left[\chihat -
  \left(\ev{\tau} - \zetahat\right) \chizerohat\right]^2}{2
  \left(\ev{\tau} - \zetahat\right)\left(1-\ev{\tau} +
  \zetahat\right)} \right\}\left[1 -
  \exp\left(-\dfrac{2\chihat\chizerohat}{1-\ev{\tau} +
    \zetahat}\right) \right],
\end{aligned}
\label{eq:PzetaphiGivenLT}
\end{equation}
where we have introduced the curvature fluctuation in unit of the
lifetime
\begin{equation}
\zetahat \equiv \dfrac{\zeta}{\rv{\Nzero}}\,.
\label{eq:zetahatdef}
\end{equation}
The Heaviside functions appearing in \cref{eq:PzetaphiGivenLT} enforce
the consistency $0<\rv{N}<\rv{\Nzero}$, which, in terms of $\zeta$,
translates to $0<\ev{\tau}-\zetahat < 1$. This is a window function
selecting subdomains for the field $\chihat$ where a given value of
$\zeta$ can be generated. These regions can be visualised by
translating the vertical axis of \cref{fig:taumean} by $\zetahat$ and
keeping only the field domains for which the shifted curves remain
within the interval $[0,1]$.

\begin{figure}
\begin{center}
  \includegraphics[width=\onefigw]{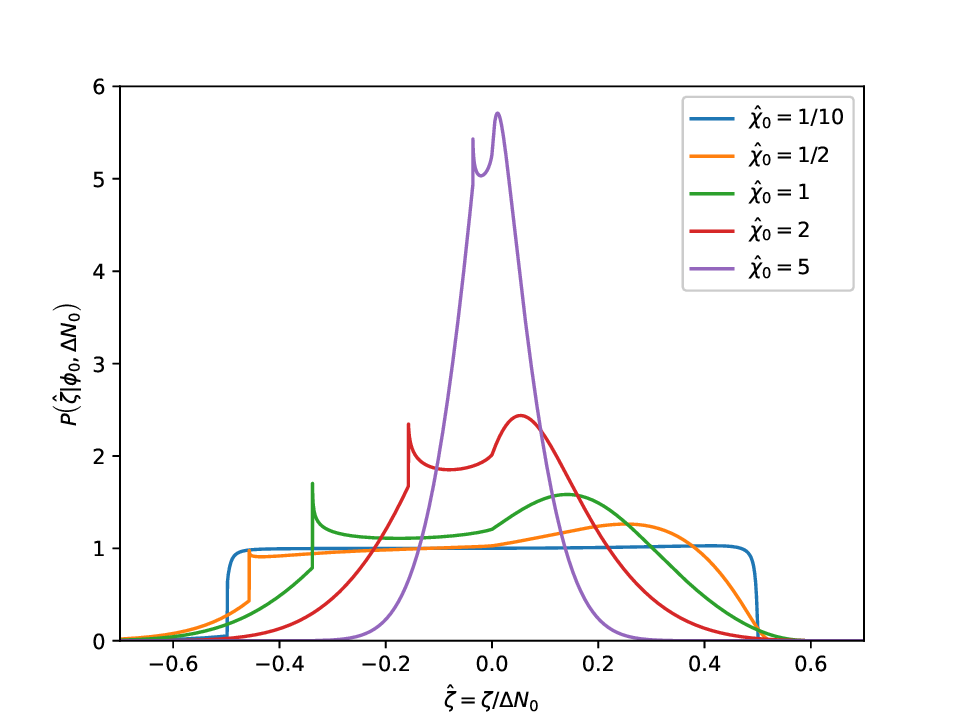}
\caption{The normalised probability distribution of the rescaled
  curvature fluctuation $\zetahat=\zeta/\rv{\Nzero}$, at given
  lifetime, plotted for various values of
  $\chizerohat=\rv{\phizero}/(G\sqrt{\rv{\Nzero}})$. For $\chizerohat
  \ll 1$, it converges to the rectangle function of
  \cref{eq:Pzetatauhalf}. In the opposite limit, $\chizerohat \gg 1$,
  it develops an asymmetric structure, due to the effect of the
  quantum wall, while becoming strongly peaked around
  $\zetahat=0$. All dependence in the lifetime $\rv{\Nzero}$ are
  absorbed in the rescaled quantities, $\zetahat$ and $\chizerohat$.}
\label{fig:Pzetahat}
\end{center}
\end{figure}

Since we are not particularly interested in the field values $\rv{\phi}$,
we can marginalise over to obtain
\begin{equation}
\Pof{\zetahat|\phizero,\rv{\Nzero}} = \int_0^{+\infty}
\Pof{\phi,\zetahat|\phizero,\rv{\Nzero}} \dd{\rv{\phi}} = G
\sqrt{\rv{\Nzero}} \int_0^{+\infty} \Pof{\phi,\zetahat|\phizero,\rv{\Nzero}} \dd{\chihat},
\label{eq:PzetaGivenLTdef}
\end{equation}
the integrand being given by \cref{eq:PzetaphiGivenLT}. Given the
functional $\ev{\tau}$ in \cref{eq:taumean}, this integral does not
seem to have an evident analytical expression. However, as we
have seen in \cref{sec:taumean}, for $\chihat \gg \chizerohat$, and
also for $\chizerohat \ll 1$, we have $\ev{\tau} \to 1/2$. In that
situation, the integral in \cref{eq:PzetaGivenLTdef} is
straightforward\footnote{The reverse Fokker-Planck
equation~\eqref{eq:FKrvsto} is conserving probability, integrating its
solution over field values gives unity.} and one gets
\begin{equation}
\Pof{\zetahat|\phizero,\rv{\Nzero}} \underset{\ev{\tau}\to 1/2}{=}
\heaviside{\dfrac{1}{2}-\zetahat} - \heaviside{-\dfrac{1}{2} - \zetahat}.
\label{eq:Pzetatauhalf}
\end{equation}
This is the rectangle function, a flat distribution which is
non-vanishing for $-\rv{\Nzero}/2 < \zeta<\rv{\Nzero}/2$. It
corresponds to the cases in which the effects coming from the quantum
wall, i.e. the end of inflation hypersurface, are negligible compared
to the quantum diffusion. For the infinite flat potential,
\cref{eq:Pzetatauhalf} shows that all values of $\zeta$ are equally
possible, but bounded by half of the lifetime.

In \cref{fig:Pzetahat}, we have plotted the exact normalised
probability distribution for $\zetahat$, as obtained by a numerical
integration of \cref{eq:PzetaGivenLTdef}. Let us notice that the
prefactor in $G\sqrt{\rv{\Nzero}}$ cancels between
\cref{eq:PzetaphiGivenLT} and \cref{eq:PzetaGivenLTdef} such that all
explicit dependence in the lifetime $\rv{\Nzero}$ disappears. The
distribution ends up being a function of $\chizerohat$ and $\zetahat$
only. The discontinuities and derivative jumps visible in
\cref{fig:Pzetahat} are not numerical artifacts, they do come from the
convolution of the Heaviside functions in \cref{eq:PzetaphiGivenLT}
within the integral of \cref{eq:PzetaGivenLTdef}. These kicks are
therefore associated with the constraints that, at given lifetime
$\rv{\Nzero}$ and field value $\rv{\phi}$, not all values of $\zeta$
can be generated. In the limit $\chizerohat \ll 1$, we recover
$\ev{\tau}=1/2$ and the distribution is given by
\cref{eq:Pzetatauhalf}. The opposite limit, $\chizerohat \gg 1$, is
quite different as the distribution develops an asymmetric structure
for small values of $\zetahat$. Notice that $\chizerohat \gg 1$ means
either $\rv{\phizero} \gg G \sqrt{\rv{\Nzero}}$ or, equivalently,
$\rv{\Nzero} \ll \rv{\phizero}^2/G^2$. As already discussed in
\cref{sec:exactsol}, this is the regime in which the field flows from
the quantum wall to the sink without much diffusion around. As such,
it is natural that the $\zetahat$ distribution becomes sharply peaked
around vanishing values. The asymmetry between positive and negative
values of $\zetahat$ can be associated with the strongly felt presence
of the wall repulsion and of the sink attraction, which are quite
different.

Finally, let us stress that the distribution of $\zetahat$, as
computed by \cref{eq:PzetaGivenLTdef}, is normalised
\begin{equation}
\int_{-\infty}^{+\infty} \Pof{\zetahat|\phizero,\rv{\Nzero}}
\dd{\zetahat} = 1,
\end{equation}
and, in spite of its asymmetry at large $\chizerohat$, we always have
\begin{equation}
\ev{\zetahat} = \int_{-\infty}^{+\infty} \zetahat \Pof{\zetahat|\phizero,\rv{\Nzero}}
\dd{\zetahat} = 0,
\end{equation}
as expected from \cref{eq:zetarv}.

\subsection{Quantum-generated curvature distribution}
\label{sec:pzeta}

This is not the end of the story as all of our reverse-time
realisations are conditioned by the lifetime $\rv{\Nzero}$. However,
we can now add-up all the lifetimes since, by definition, one has
\begin{equation}
\Pof{\zeta|\phizero} = \int_0^{+\infty}
\Pof{\zeta,\rv{\Nzero}|\phizero} \dd{\rv{\Nzero}},
\label{eq:Pzetadef}
\end{equation}
where, from the product rule,
\begin{equation}
\Pof{\zeta,\rv{\Nzero}|\phizero} = \Pof{\zeta|\phizero,\rv{\Nzero}} \Pof{\rv{\Nzero}|\phizero}.
\end{equation}
The last term of this equation is the probability distribution of the
lifetimes, $\Plt{\rv{\Nzero}|\phizero}$ of \cref{eq:Pltsto}. Moreover,
we also have
\begin{equation}
\Pof{\zeta|\phizero,\rv{\Nzero}} = \dfrac{1}{\rv{\Nzero}} \Pof{\zetahat|\phizero,\rv{\Nzero}},
\end{equation}
from which one gets
\begin{equation}
\Pof{\zeta|\phizero} = \dfrac{\chizero}{\sqrt{2\pi}} \int_0^{+\infty}
\Pof{\zetahat|\phizero,\rv{\Nzero}} \dfrac{e^{-\frac{\chizero^2}{2\rv{\Nzero}}}}{\rv{\Nzero}^{5/2}}\dd{\rv{\Nzero}}.
\label{eq:Pzetaint}
\end{equation}
One should pay attention to the dependence
$\zetahat=\zeta/\rv{\Nzero}$ in the previous integral. Moreover, the
window function in \cref{eq:PzetaphiGivenLT} enforces that the
integrand vanishes for $\abs{\zeta}>\rv{\Nzero}$ and the lower bound
can equally be chosen at $\abs{\zeta}$.

We can simplify the previous expression by changing the integration
variable from $\rv{\Nzero}$ to
$\chizerohat=\chizero/\sqrt{\rv{\Nzero}}$, at fixed $\chizero$, to
obtain another exact expression
\begin{equation}
\Pof{\zeta|\phizero} = \dfrac{1}{\chizero^2}  \sqrt{\dfrac{2}{\pi}} \int_0^{\frac{\chizero}{\sqrt{\abs{\zeta}}}}
\Pof{\zetahat=\dfrac{\zeta}{\chizero^2}\chizerohat^2|\phizero,\rv{\Nzero}}
\chizerohat^2 e^{-\frac{\chizerohat^2}{2}} \dd{\chizerohat},
\label{eq:Pzetafinal}
\end{equation}
where we recap that there is no explicit dependence in $\rv{\Nzero}$
within the function $\Pname(\zetahat|\phizero,\rv{\Nzero})$.  As a
result, $\chizero^2 \Pof{\zeta|\phizero}$ is a function of
$\zeta/\chizero^2$ only.

We have numerically integrated \cref{eq:Pzetafinal}, using the
numerical solutions of \cref{sec:PzetaGivenLT}. It is, however,
instructive to calculate analytically this integral in the large
lifetime limit, namely assuming that $\ev{\tau}=1/2$. In this case,
the distribution for $\zetahat$ is the rectangle function and plugging
\cref{eq:Pzetatauhalf} into \cref{eq:Pzetaint} yields
\begin{equation}
\chizero^2 \Pof{\zeta|\phizero} \underset{\ev{\tau}\to/1/2}{=}
  \erf\left(\dfrac{\chizero}{2\sqrt{\abs{\zeta}}}\right) -
  \dfrac{\chizero}{\sqrt{\abs{\zeta}}}
  \dfrac{e^{-\frac{\chizero^2}{4\abs{\zeta}}}}{\sqrt{\pi}}\,.
\label{eq:Pzetainf}
\end{equation}

\begin{figure}
\begin{center}
  \includegraphics[width=\onefigw]{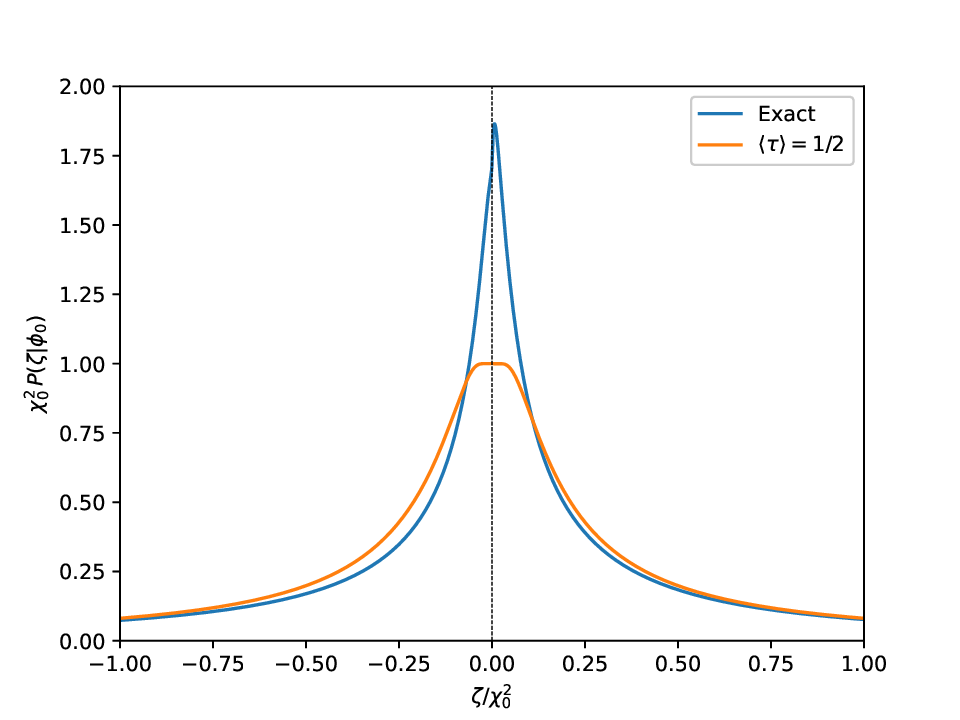}
  \includegraphics[width=\onefigw]{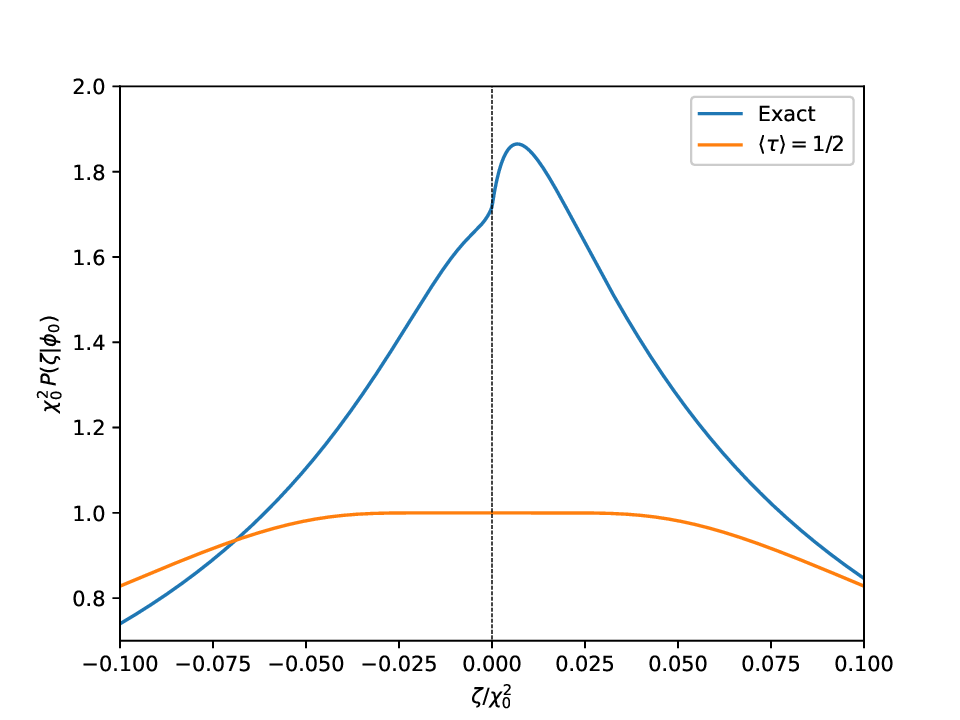}
\caption{The rescaled normalised probability distribution $\chizero^2
  \Pof{\zeta|\phizero}$ (blue curve) of the curvature fluctuation
  $\zeta$, plotted as a function of $\zeta/\chizero^2$. It is slightly
  skewed towards positive values due to the presence of the quantum
  wall. Its tails are heavy and decay as $1/\abs{\zeta}^{3/2}$. The
  lower panel is a zoom around the small values of $\zeta/\chizero^2$,
  the mode being given by \cref{eq:Pzetamode}. The orange curve is
  \cref{eq:Pzetainf}, an analytical approximation obtained in the very
  large lifetime limit for which $\ev{\tau}=1/2$.}
\label{fig:Pzeta}
\end{center}
\end{figure}

In \cref{fig:Pzeta}, we have plotted both the exact numerical
integration of \cref{eq:Pzetafinal} and the infinite lifetime
approximation of \cref{eq:Pzetainf}. The exact expression matches the
approximated one in the tails, showing that large curvature
fluctuations are mostly induced by the large lifetime realisations. As
can be seen by carefully expanding \cref{eq:Pzetainf} in the large
$\zeta/\chizero^2$ limit, the tails are the same as a Levy
distribution, they decay as $1/\abs{\zeta}^{3/2}$. This is a
normalisable distribution, with, however, an infinite variance. There
is also a change of behaviour at small values of $\zeta/\chizero^2$,
typically for $\abs{\zeta}/\chizero^2<1/10$, the position of the
inflection point of \cref{eq:Pzetainf}. The distribution becomes
narrower, it is skewed towards positive curvatures and the mode is at a
value that we numerically determine to be
\begin{equation}
  \dfrac{\zeta_{\umode}}{\chizero^2} \simeq 6.9\times 10^{-3}.
\label{eq:Pzetamode}
\end{equation}

The method we have presented is not restricted to $\zeta$. Various
other quantities of interest could be obtained using the reverse-time
formalism. The main idea is to first account for the stochasticity at
given lifetime, and, then, to marginalise over all possible lifetimes,
including the infinite limit. As a motivated example, we show in
\cref{sec:Pbw} how to derive the so-called backward distribution of
field values along these lines.

To conclude this section, let us remark that the marginalisation of
\cref{eq:Pzetadef} includes $\rv{\Nzero}\to \infty$. Since the
survival probability of \cref{eq:survival} vanishes in the limit of
infinite lifetime, for the semi-infinite flat potential, our
reverse-time formalism for stochastic inflation allows for the
asymptotic inclusion of eternally inflating domains, and so does
\cref{eq:Pzetafinal}. In more complex situations, as for instance if
stochastic inflation occurs with more than one field, the survival
probability may not vanish. Still, the limit $\rv{\Nzero}\to \infty$
of the time-reversed stochastic processes could be taken and our
formalism could be viewed as a natural choice for a measure. This
measure would include realisations in which stochastic inflation ended
after an infinite lifetime while discarding the never-ending inflating
cases.

\section{Conclusion}
\label{sec:conc}

In this work, we have shown that stochastic inflation could be dealt
with in a time-reversed approach. The coarse-grained field emerges
from the quantum wall and evolves towards a sink that is necessarily
reached at a time matching the lifetime of the forward process. The
associated probability distribution of the field values has been shown
to follow a reverse Fokker-Planck equation which exhibits a new
friction term thereby ensuring the correspondence with the forward
process. As discussed in \cref{sec:trfk}, in a forward-only picture,
the time-reversal may also be interpreted as a conditioning, and
partitioning, of the forward realisations by their lifetimes, the
friction term emerging from a Doob's $h$-transform.

In the simple case where the potential is exactly flat over a
semi-infinite domain, time-reversed quantum diffusion ends up being
exactly solvable and we have obtained in \cref{eq:solrv} the field
distribution $\rv{N}$ {\efolds} before the end of stochastic
inflation, conditioned by the lifetime $\rv{\Nzero}$. Combined with a
``time-reversed'' version of the stochastic $\deltaN$-formalism, this
has allowed us to derive a new and finite formula for the probability
distribution of the quantum-generated curvature fluctuations in
\cref{eq:Pzetafinal}, which is plotted in \cref{fig:Pzeta}. All these
quantities are finite, possibly universal, and one may ask why
reversing time appears to cure the divergences of the forward
formalism for the semi-infinite flat potential. In fact, it is not
only reversing time, but rather, by doing so, one enforces a
conditioning to $\rv{\Nzero}$ first. In other words, taking average
values and marginalising over $\rv{\Nzero}$ do not commute. In the
forward stochastic $\deltaN$-formalism, one tries to average first,
and this leads to divergences. Here, at given lifetime $\rv{\Nzero}$,
averaging is necessarily finite and marginalising afterwards gives the
finite distribution of \cref{eq:Pzetafinal}. Notice, however, that
this distribution has the same heavy tails as a Levy's: mean and
variance are actually undefined, as we should expect from the very
same arguments.

In qualitative terms, our result also shows that nothing really bad
occurs within the flat semi-infinite potential. The distribution of
$\zeta$ is significant only for values $\abs{\zeta} \simeq
\chizero^2=\rv{\phizero}^2/G$, i.e., determined only by the initial
value of the coarse grained field and the quantum diffusion
coefficient (the Hubble parameter during inflation). On general
grounds, we have demonstrated that focusing only on pathological
stochastic trajectories (e.g. the one that diffuse forever) can
provide misleading results. Indeed, the well-definiteness of the
$\zeta$-distribution builds upon a careful resummation of \emph{all}
possible trajectories.

Even though we have only worked out the semi-infinite flat potential,
the time-reversed stochastic formalism could be applied to any
stochastic inflation models in any potentials. However, the friction
term of the reverse Fokker-Planck equation can be quite complicated
and finding exact, or numerical, solutions for the reverse probability
distribution is, a priori, more difficult than for the forward
process. As already discussed, the advantage of the time-reversal lies
in the conditioning by the lifetimes, and this is a powerful manner to
cure some divergences. For instance, it would be interesting to apply
it to the bounded quantum-well, which exhibits divergences typical of
a phase transition when the field excursion of the well exceeds some
critical value. Let us also stress that, in the time-reverse
formalism, all probabilities are necessarily expressed in number of
{\efolds} in reference to the end-of-inflation hypersurface. As such,
they are almost ready-to-use from an observational point of view.

Another possible applications of the time-reverse formalism concerns
eternal inflation. The formalism indeed includes eternally inflating
domains corresponding to the limit $\rv{\Nzero} \to \infty$. However,
as we have argued, in multifield stochastic inflation, this limit may
actually be missing an infinite number of eternally inflating regions,
and, as such, it could be viewed as a natural regulator keeping
domains in which stochastic inflation ends after an infinite lifetime
while dropping others in which it is never ending.

On more fundamental grounds, stochastic quantization is an approach to
quantum mechanics in which Hilbert spaces are replaced by random
processes and time-reversal plays in a major
role~\cite{1966PhRv..150.1079N}. It would be interesting to see if one
could not start from stochastic quantization from the very beginning,
even before performing the coarse-graining of quantum operators that
allows us to deal with stochastic inflation. In this manner, one could
possibly account for the entanglement effects which are here discarded
by the coarse-graining. On the gravity side, it is clear that as soon
as $\zeta \gg 1$, gravitational collapse should strongly affect the
distribution of curvature fluctuations, and, it could be interesting
to extend the formalism to positive curvature
spaces~\cite{Handley:2019anl, Letey:2022hdp, Grain:2020wro}.

\section*{Acknowledgements}
We would like thank C.~Animali, P.~Auclair, E.~Tomberg and V.~Vennin
for enlightening discussions and insightful comments that have helped
us to improve the quality of the manuscript. This work is supported by
the ESA Belgian Federal PRODEX Grants $\mathrm{N^{\circ}} 4000143201$
and $\mathrm{N^{\circ}} 4000144768$. B.~B. is publishing in the
quality of ASPIRANT Research Fellow of the FNRS.

\appendix

\section{Exact solution from Girsanov's method}
\label{sec:girsanov}
In order to solve the reverse Fokker-Planck \cref{eq:FKrvsto} with the
friction term of \cref{eq:Frvsto}, we can use the Girsanov's
theorem~\cite{Sarkka_Solin_2019}, which quantifies the effect of a
change of probabilistic measure in stochastic differential
equations. Let us consider the process
\begin{equation}
\dd{\Xrv} = \Frv(\Xrv,t) \dd{t} + \dd{W(t)}, \qquad \Xrv(0) = 0,
\label{eq:Xrvito}
\end{equation}
where
\begin{equation}
\Frv(x,t) = -\dfrac{x}{\tzero -t} + \dfrac{\chizero}{\tzero-t}
\coth\left(\dfrac{\chizero x}{\tzero - t}\right),
\label{eq:Fxt}
\end{equation}
with $t<\tzero$. The Girsanov's theorem implies that for some regular enough functional $h(x)$
\begin{equation}
\mathbb{E}\left[h\left(\left\{X\right\}\right)\right] = \mathbb{E}\left[Z(t) h\left(\left\{W\right\}\right) \right],
\label{eq:esperance}
\end{equation}
where the braces refer to the ensemble of all realisations and $Z(t)$ is given by the stochastic integral
\begin{equation}
Z(t) = \exp{\int_0^t \Frv \left[W(u),u\right] \dd{W(u)} -
  \dfrac{1}{2} \int_0^t \Frv^2\left[W(u),u\right] \dd{u}}.
\label{eq:girsanov}
\end{equation}
Hence, a change of measure can be performed in order to recover a driftless Brownian motion 
(with known probability density), and, if the expression for $Z(t)$ is sufficiently simple,
a closed-form of the probability density can be derived.

Following Ref.~\cite{Mazzolo2024}, we consider the function
\begin{equation}
f(x,t) = \ln\left[\sinh\left(\dfrac{\chizero
    x}{\tzero - t}\right) \right] - \dfrac{x^2}{2\left(\tzero-t\right)}\,,
\label{eq:fdef}
\end{equation}
which, from the It\^o's formula, verifies
\begin{equation}
\dd{f\left[W(u),u\right]} = \eval{\left( \pdv{f}{t} + \dfrac{1}{2}
\pdv[2]{f}{x} \right)}_{W(u),u}\dd{u} + \eval{\pdv{f}{x}}_{W(u),u} \dd{W(u)},
\end{equation}
or
\begin{equation}
  \begin{aligned}
\dd{f\left[W(u),u\right]} & = \Bigg[
-\dfrac{W^2}{2\left(\tzero-u\right)^2} + \dfrac{\chizero
  W}{\left(\tzero-u\right)^2} \coth\left(\dfrac{\chizero
    W}{\tzero-u}\right) - \dfrac{\chizero^2}{2
  \left(\tzero-u\right)^2} \coth^2\left(\dfrac{\chizero
    W}{\tzero-u}\right) \\ & -\dfrac{1}{2\left(\tzero-u\right)} +
\dfrac{\chizero^2}{2\left(\tzero-u\right)^2}\Bigg] \dd{u} +
\left[-\dfrac{W}{\tzero-u} + \dfrac{\chizero}{\tzero-u}
\coth\left(\dfrac{\chizero W}{\tzero-u}\right)\right]\dd{W}.
  \end{aligned}
\label{eq:df}
\end{equation}
In this expression $W$ stands for $W(u)$ and the term multiplying the
noise is exactly $\Frv[W(u),u]$. Thus, we can derive an exact
expression for the stochastic integral
\begin{equation}
  \begin{aligned}
\int_0^t \Frv(W,u) \dd{W} & = f[W(t),t] -f[W(0),0] + \int_0^t\Bigg[
\dfrac{W^2}{2\left(\tzero-u\right)^2} - \dfrac{\chizero
  W}{\left(\tzero-u\right)^2} \coth\left(\dfrac{\chizero
    W}{\tzero-u}\right) \\ & + \dfrac{\chizero^2}{2
  \left(\tzero-u\right)^2} \coth^2\left(\dfrac{\chizero
    W}{\tzero-u}\right) + \dfrac{1}{2\left(\tzero-u\right)} -
\dfrac{\chizero^2}{2\left(\tzero-u\right)^2}\Bigg] \dd{u}.
  \end{aligned}
\end{equation}
This integral is the first term in the argument of the exponential in
\cref{eq:girsanov}. The second term is an integral over $\Frv^2$, and,
from \cref{eq:Fxt}, one gets
\begin{equation}
  \begin{aligned}
\int_0^t \Frv[W(u),u] \dd{W(u)} & - \dfrac{1}{2} \int_0^t
\Frv^2\left[W(u),u\right]\dd{u} = f[W(t),t] -f[W(0),0] \\ & +
\ln\left(\sqrt{\dfrac{\tzero}{\tzero-t}}\right) - \dfrac{\chizero^2
  t}{2\tzero\left(\tzero-t\right)}\,.
  \end{aligned}
\end{equation}
Plugging this expression into \cref{eq:girsanov}, one obtains
\begin{equation}
Z(t) = \dfrac{\sinh\left[\dfrac{\chizero
      W(t)}{\tzero-t}\right]}{\sinh\left[\dfrac{\chizero
      W(0)}{\tzero}\right]} \, e^{-\frac{W^2(t)}{2\left(\tzero-t\right)}
  + \frac{W^2(0)}{2\tzero}} \sqrt{\dfrac{\tzero}{\tzero-t}} \,
e^{-\frac{\chizero^2 t}{2 \tzero \left(\tzero-t\right)}}.
\label{eq:zt}
\end{equation}
For our purpose, we can replace the pure Brownian motion $W(t)$ by a
shifted Brownian motion $X$, starting at $X(0)=\xzero$, in presence of
an absorbing boundary at $x=0$. The transition probability for such a
process has been derived in \cref{eq:Pfwsto} and reads
\begin{equation}
\Pof{x,t|\xzero,0} = \dfrac{1}{\sqrt{2\pi t}}
\left[e^{-\frac{\left(x-\xzero\right)^2}{2t}} -
    e^{-\frac{\left(x+\xzero\right)^2}{2t}} \right].
\end{equation}
From \cref{eq:esperance,eq:zt}, the probability density associated with
the process $\Xrv$ of \cref{eq:Xrvito}, now starting at
$\Xrv(0)=\xzero$, is thus given by $\Prv{x,t|\xzero,0} =
Z(t)\Pof{x,t|\xzero,0}$, i.e.,
\begin{equation}
\Prv{x,t|\xzero,0} = \sqrt{\dfrac{\tzero}{2 \pi
    t\left(\tzero-t\right)}} e^{-\frac{\chizero^2
    t}{2\tzero(\tzero-t)}} \left[
  e^{-\frac{\left[\tzero x-(\tzero-t)\xzero\right]^2}{2t\tzero(\tzero-t)}} -
  e^{-\frac{\left[\tzero
        x+(\tzero-t)\xzero\right]^2}{2t\tzero(\tzero-t)}}  \right]
  \dfrac{\sinh\left(\dfrac{\chizero
      x}{\tzero-t}\right)}{\sinh\left(\dfrac{\chizero \xzero}{\tzero}\right)}\,. 
\label{eq:Prvxzero}
\end{equation}
However, as discussed in \cref{sec:revsto}, for time-reversed
stochastic inflation, the reverse process has to start exactly on the
boundary, namely for $\xzero=0$. Taking the limit $\xzero \to 0$ in
\cref{eq:Prvxzero} finally gives
\begin{equation}
\Prv{x,t|0,0} = \sqrt{\dfrac{2}{\pi}}\sqrt{\dfrac{\tzero^3}{t^3\left(\tzero - t
    \right)}} \sinh\left(\dfrac{\chizero x}{\tzero-t}\right)
\dfrac{x}{\chizero} e^{-\frac{\tzero^2 x^2 + \chizero^2 t^2}{2 t
    \tzero(\tzero -t)}},
\end{equation}
which solves \cref{eq:FKrvsto}, with a friction term given by
\cref{eq:Frvsto} and with the initial and boundary conditions of
\cref{eq:Prvibc}.

\section{Exponentials of complicated arguments with powers}
\label{sec:Kintegral}

In this section, we derive an exact analytical expression for the
integrals of the form
\begin{equation}
K(\beta,\gamma) = \int_0^{+\infty} \dfrac{e^{-\frac{\beta}{z} - \gamma
    z}}{\sqrt{z} \left(1+z\right)} \dd{z},
\end{equation}
$\beta$ and $\gamma$ being positive numbers. We first split the
integral into 
\begin{equation}
K(\beta,\gamma) = \int_0^1\dfrac{e^{-\frac{\beta}{z} - \gamma
    z}}{\sqrt{z} \left(1+z\right)} \dd{z} + \int_1^{+\infty}\dfrac{e^{-\frac{\beta}{z} - \gamma
    z}}{\sqrt{z} \left(1+z\right)} \dd{z},
\end{equation}
and remark that, changing variable $z\to 1/z$ in the second integral
gives back the first, up to a permutation between $\beta$ and $\gamma$. As such,
\begin{equation}
K(\beta,\gamma) = J(\beta,\gamma) + J(\gamma,\beta),
\end{equation}
where
\begin{equation}
J(\beta,\gamma) \equiv \int_0^1 \dfrac{e^{-\frac{\beta}{z} - \gamma
    z}}{\sqrt{z} \left(1+z\right)} \dd{z}.
\end{equation}
Because this integral is over a domain in which $\abs{z}<1$, we can
Taylor expand the term in $1/(1+z)$ and
\begin{equation}
J(\beta,\gamma) = \sum_{n=0}^{\infty} (-1)^n I_n(\beta,\gamma), 
\end{equation}
where
\begin{equation}
I_n(\beta,\gamma) \equiv \int_0^1 z^{n-1/2} e^{-\frac{\beta}{z} - \gamma
    z} \dd{z}.
\end{equation}
The family of integrals $I_n$ are all generated by differentiation
with respect to $\beta$ or $\gamma$
\begin{equation}
\pdv{I_n}{\beta} = -I_{n-1}, \qquad \pdv{I_n}{\gamma} = -I_{n+1},
\label{eq:pdvIn}
\end{equation}
and this allows us to write a partial differential equation system
satisfied by $K(\beta,\gamma)$. One gets
\begin{equation}
\left\{
  \begin{aligned}
\pdv{K(\beta,\gamma)}{\beta} & = K(\beta,\gamma) - \left[I_0(\gamma,\beta) +
I_{-1}(\beta,\gamma)\right],\\
\pdv{K(\beta,\gamma)}{\gamma} & = K(\beta,\gamma) -
\left[I_0(\beta,\gamma) + I_{-1}(\gamma,\beta) \right].
  \end{aligned}
  \right.
\label{eq:Kpde}
\end{equation}
The integral $I_0(\beta,\gamma)$ is known~\cite{gradshteyn2007} and
$I_{-1}(\beta,\gamma)$ can be obtained by differentiation using \cref{eq:pdvIn}. One can
also split the integral in \cref{eq:GR_p3.475eq15} in the two domains $[0,1]$
and $[1,+\infty[$ to show that
\begin{equation}
I_0(\gamma,\beta) + I_{-1}(\beta,\gamma) = \sqrt{\dfrac{\pi}{\beta}}
e^{-2\sqrt{\beta\gamma}}, \qquad I_0(\beta,\gamma) +
I_{-1}(\gamma,\beta) =  \sqrt{\dfrac{\pi}{\gamma}}
e^{-2\sqrt{\beta\gamma}}.
\end{equation}
Plugging these source terms into \cref{eq:Kpde}, the system can be
straightforwardly integrated. The unknown integration constant is fixed
by enforcing the limit $K(\infty,\infty)= 0$ and one finally gets
\begin{equation}
K(\beta,\gamma) = \pi e^{\beta+\gamma} \erfc\left(\sqrt{\beta}+\sqrt{\gamma}\right).
\end{equation}

\section{Backward probability distribution}
\label{sec:Pbw}

\begin{figure}
\begin{center}
  \includegraphics[width=\onefigw]{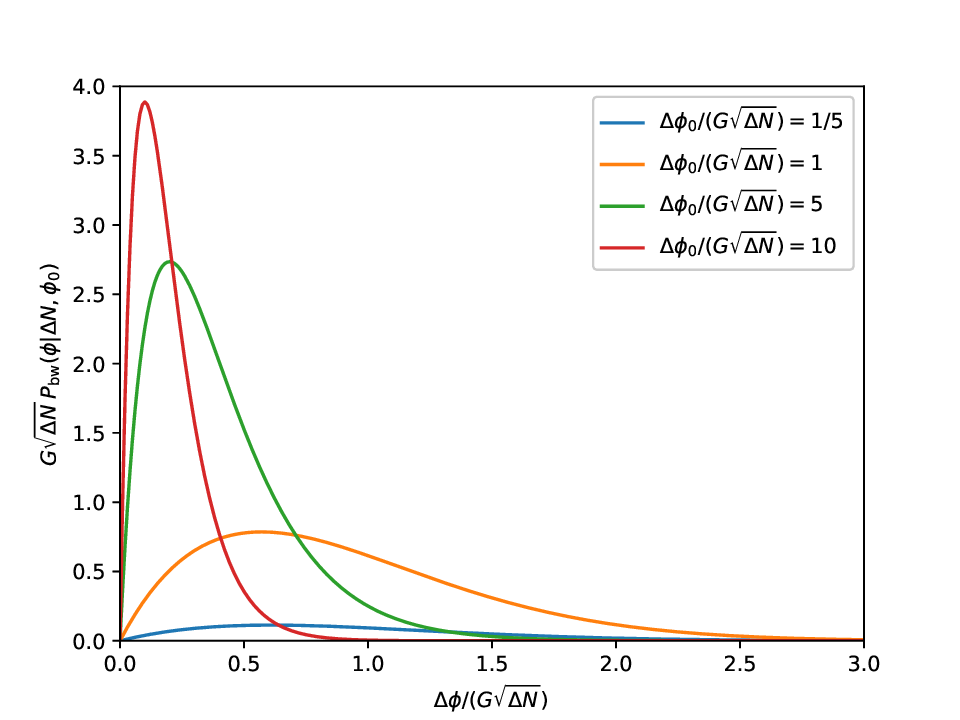}
\caption{The rescaled backward probability distribution
  $G\sqrt{\rv{N}}\,\Pbw{\phi|\rv{N},\phizero}$, after marginalisation
  over all lifetimes, as a function of $\chi/\sqrt{\rv{N}}$ and
  plotted for various values of
  $\chizero/\sqrt{\rv{N}}$.}
\label{fig:Pbw}
\end{center}
\end{figure}

Another quantity of interest is the so-called backward distribution of
the field values, given some {\efolds} before the end of stochastic
inflation~\cite{Ando:2020fjm}. In the reverse-time approach, we are
now interested in taking a slice of \cref{fig:rvprob} at constant
$\rv{N}$ and consider the stochasticity of $\rv{\phi}$, i.e.,
$\Pof{\phi|\rv{N},\phizero,\rv{\Nzero}} \propto
\Prv{\phi,\rv{N}|\phizero,\rv{\Nzero}}$, which is conditioned by the
lifetime $\rv{\Nzero}$. In order to marginalise over the lifetimes, we
need the joint probability
\begin{equation}
  \Pof{\phi,\rv{\Nzero}|\rv{N},\phizero} =
  \Pof{\phi|\rv{N},\phizero,\rv{\Nzero}} \Pof{\rv{\Nzero}|\rv{N},\phizero}.
\label{eq:Pbwjt}
\end{equation}
From the Bayes's theorem, one has
\begin{equation}
  \Pof{\rv{\Nzero}|\rv{N},\phizero} =
  \dfrac{\Plt{\rv{\Nzero}|\phizero} \Pof{\rv{N}|\phizero,\rv{\Nzero}}}{\Pof{\rv{N}|\phizero}}\,,
\label{eq:Pltrv}
\end{equation}
where the denominator is a normalisation factor. From
\cref{eq:Pbwjt,eq:Pltrv}, one obtains
\begin{equation}
\Pof{\phi,\rv{\Nzero}|\rv{N},\phizero} =
\Prv{\phi,\rv{N}|\phizero,\rv{\Nzero}} \dfrac{
  \Plt{\rv{\Nzero}|\phizero} \Pof{\rv{N}|\phizero,\rv{\Nzero}}
}{\Pof{\rv{N}|\phizero}}\,.
\end{equation}
One has to determine $\Pof{\rv{N}|\phizero,\rv{\Nzero}}$, which, up to
a normalisation constant, can be obtained by integrating
\cref{eq:solsimple} over all field values. The mass conservation
property of the reverse Fokker-Planck equation ensures that
\begin{equation}
\Pof{\rv{N}|\phizero,\rv{\Nzero}} \propto \int_0^{+\infty}
\Prv{\phi,\rv{N}|\phizero,\rv{\Nzero}} \dd{\rv{\phi}} = 1,\quad \forall \rv{N}\in]0,\rv{\Nzero}[.
\label{eq:normN}
\end{equation}
This is therefore a flat distribution, non-vanishing for
$0<\rv{N}<\rv{\Nzero}$. In order to interpret it as a probability
distribution for $\rv{N}$, it needs to be normalised and one has
\begin{equation}
\Pof{\rv{N}|\phizero,\rv{\Nzero}} =
\dfrac{\heaviside{\rv{N}}-\heaviside{\rv{N}-\rv{\Nzero}}}{\rv{\Nzero}}\,.
\label{eq:PrvN}
\end{equation}
Plugging this expression into \cref{eq:Pltrv}, the backward probability distribution finally reads
\begin{equation}
\Pbw{\phi|\rv{N},\phizero} =
\dfrac{1}{\Pof{\rv{N}|\phizero}}\int_{\rv{N}}^{+\infty}
\dfrac{\Prv{\phi,\rv{N}|\phizero,\rv{\Nzero}}}{\rv{\Nzero}} \Plt{\rv{\Nzero}|\phizero}\dd{\rv{\Nzero}}.
\label{eq:Pbwdef}
\end{equation}
From \cref{eq:Pltsto,eq:solsimple}, one obtains, after some algebra,
\begin{equation}
  \begin{aligned}
\Pbw{\phi|\rv{N},\phizero} & = \dfrac{1}{G \sqrt{\rv{N}}}\dfrac{\chi}{2 \sqrt{\rv{N}}} \dfrac{
e^{-\frac{\chi \chizero}{\rv{N}}}
\left[1-\erf{\left(\dfrac{\chi-\chizero}{\sqrt{2\rv{N}}}\right)}\right]
- e^{\frac{\chi\chizero}{\rv{N}}} \left[1-
  \erf{\left(\dfrac{\chi+\chizero}{\sqrt{2\rv{N}}}\right)}
  \right]}{\dfrac{\rv{N}}{\chizero^2}\erf{\left(\dfrac{\chizero}{\sqrt{2\rv{N}}}\right)}
-\sqrt{\dfrac{2}{\pi}}
\dfrac{\sqrt{\rv{N}}}{\chizero}e^{-\frac{\chizero^2}{2\rv{N}}} } .
  \end{aligned}
\label{eq:Pbw}
\end{equation}
This distribution, multiplied by $G\sqrt{\rv{N}}$, has been
represented in \cref{fig:Pbw} for various values of
$\chizero/\sqrt{\rv{N}}$. For $\rv{N} \ll \chizero^2$, it is sharply
peaked close to the quantum wall, i.e., the end of the stochastic
inflation hypersurface.

\bibliographystyle{JHEP}

\bibliography{references}

\end{document}